\documentclass{ws-rv9x6}
\usepackage{subfigure}     
\usepackage{ws-rv-van}     
\makeindex
\begin{document}

\chapter[Using World Scientific's Review Volume Document Style]
{Thomas-Fermi studies of pairing in inhomogeneous systems: nuclear and cold atom
systems at overflow \\}\label{ra_ch1}

\author[F. Author and S. Author]{Peter Schuck$^{1,2}$ and Xavier Vi\~nas$^{3}$} 

\address{$^{1}$Institut de Physique Nucl\'eaire, IN2P3-CNRS, Universit\'e 
Paris-Sud,
\\F-91406 Orsay-C\'edex, France \\
$^{2}$ Laboratoire de Physique et Mod\'elisation des Milieux Condens\'es,
CNRS and Universit\'e Joseph Fourier, 25 Avenue des Martyrs, Bo\^{i}te Postale 
166,
F-38042 Grenoble Cedex 9, France\\
$^{3}$Departament d'Estructura i Constituents de la Mat\`eria
and Institut de Ci\`encies del Cosmos, Facultat de F\'{\i}sica,
Universitat de Barcelona, Diagonal {\sl 647}, {\sl E-08028} Barcelona,
Spain \\
schuck@ipno.in2p3.fr}

\begin{abstract}
 A novel Thomas-Fermi (TF)
approach to inhomogeneous superfluid Fermi-systems is presented and shown that
it works well also in cases where the Local Density Approximation (LDA) 
breaks down. The 
novelty lies in the fact that the semiclassical approximation is applied to 
the pairing matrix elements not implying a local version of the chemical 
potential as with LDA.
Applications will be given to the generic fact that if a
fermionic superfluid in the BCS regime overflows from a narrow container
into a much wider one,
pairing is substantially reduced at the overflow point. Two examples
pertinent to the physics of the outer crust of neutron stars and superfluid
fermionic atoms in traps will be presented. The TF results will be compared 
to quantal and LDA ones.
\end{abstract}

\body

\section{Introduction}

The quantal treatment of pairing in inhomogeneous systems is a notoriously 
difficult problem. This is especially true for systems containing a large 
number $N$ of particles as it is usually the case for cold atoms in traps 
($N \sim 10^6$) \cite{pit03} or even for smaller systems if they are deformed as can 
happen for nuclei. Semiclassical approaches may 
be very helpful in such cases. The simple and very well known Local Density 
Approximation (LDA) \cite{kuch89} is not always applicable because for its validity the 
condition that the size of the Cooper pair (coherence length), $\xi$,  
must be smaller than a typical length $l$ over which the mean field potential 
is varying ($l$ is, e.g., the oscillator length in the case of a harmonic 
potential) is not always fullfilled. We here, therefore, will apply the TF 
approximation directly to the pairing matrix elements whose evaluation only 
requires the usual TF condition that the wave lengths involved must be smaller 
than $l$ \cite{vin03} . We will show that, indeed, our approach also works for 
cases 
where $\xi$ is larger than $l$ where the LDA fails. We will demonstrate 
this for the BCS approach 
in this paper. 

\begin{figure} \begin{center}
\includegraphics[height=5.5cm,angle=-90]{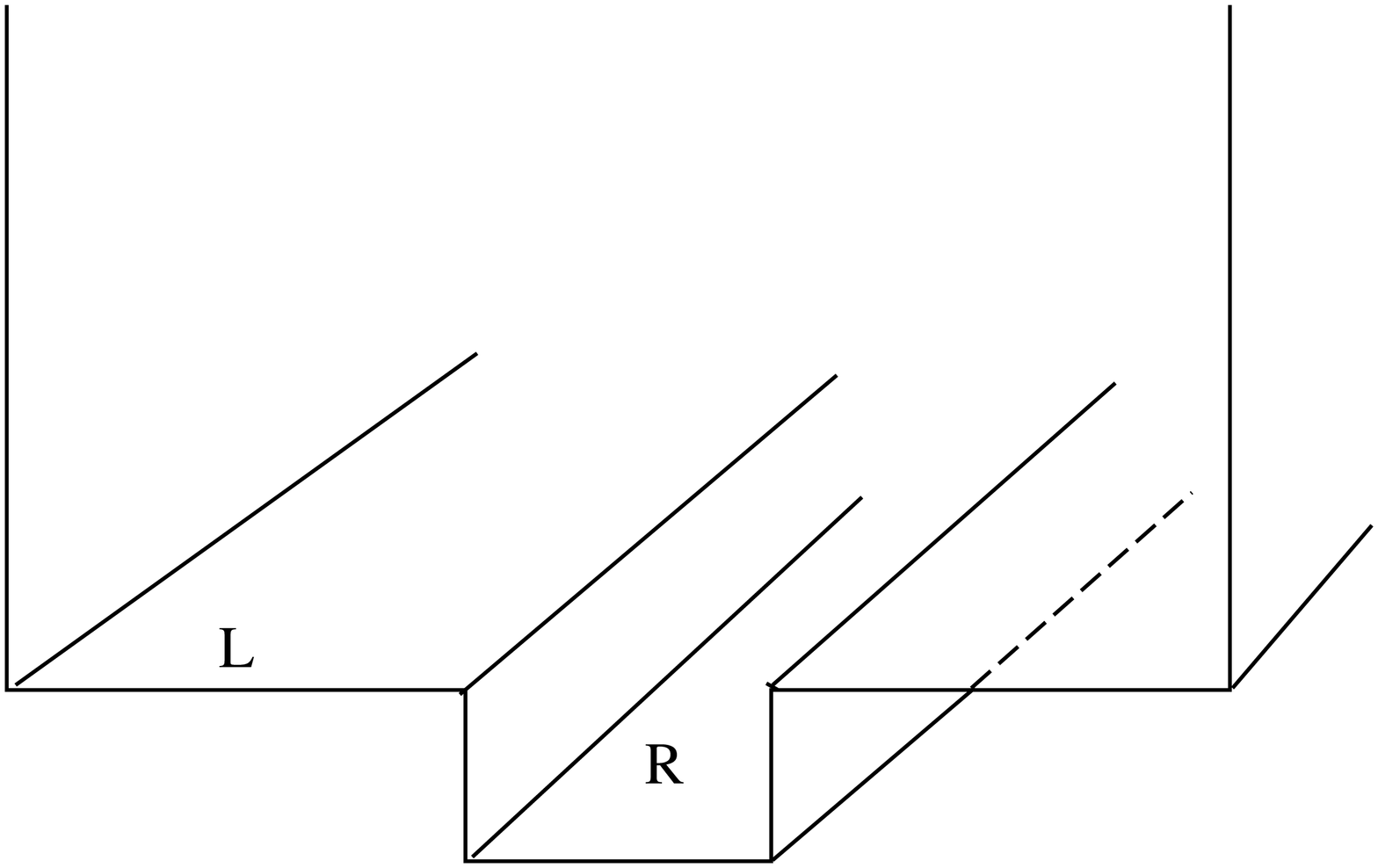}
\includegraphics[height=5.5cm,angle=-90]{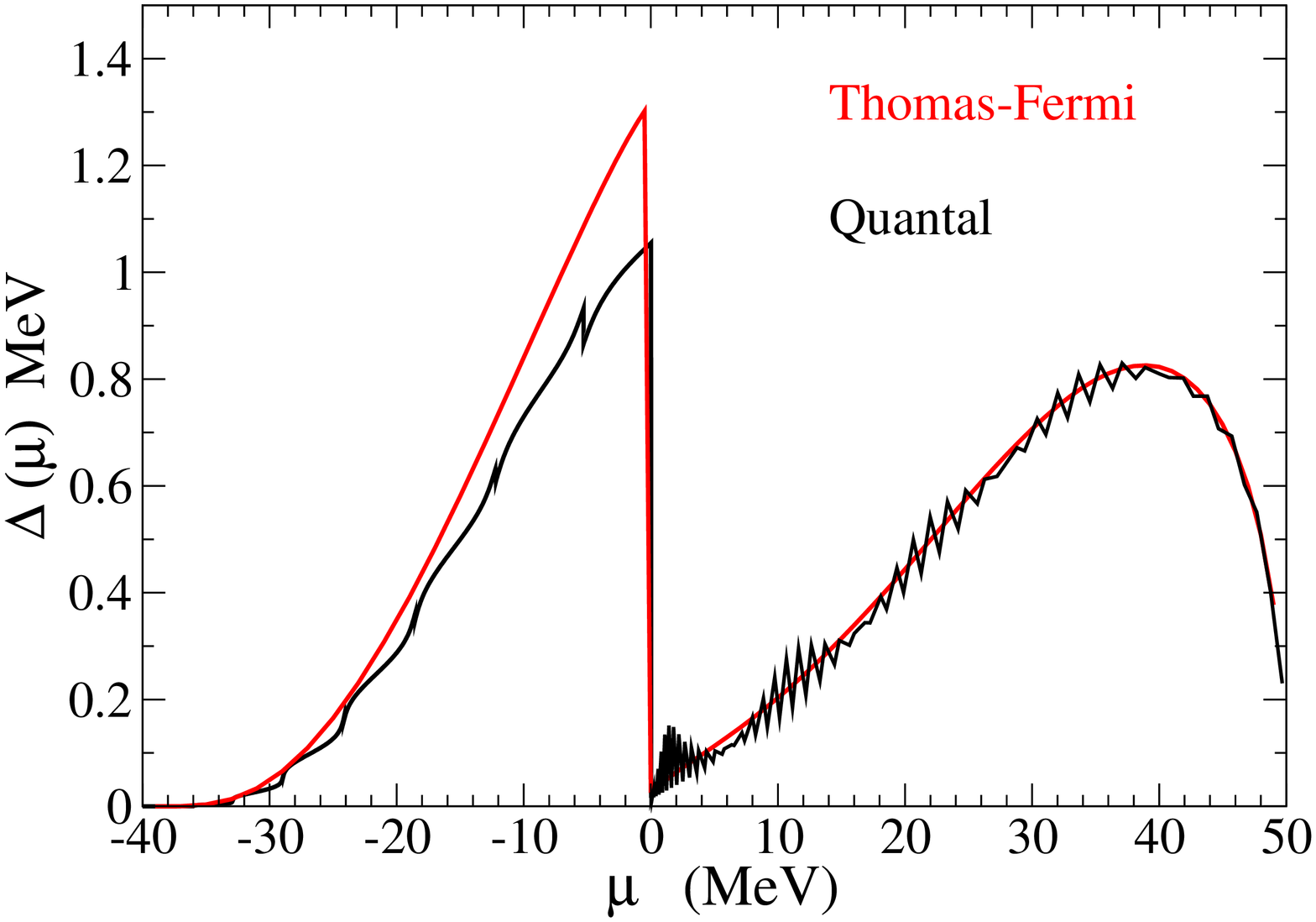}
\caption{
\label{slab1} Left: Schematic view of the slab  used in this work with a
perspective view of the potential which is
translationally invariant in x, y direction.
Right: Quantal and TF pairing gaps in
the slab geometry as a
function of the chemical potential.}
\end{center}
\end{figure}

The physical systems we are interested in concern cold atoms in traps and 
nuclei, both in so called overflow or drip configurations. For the latter 
overflow or drip means that there is such a large neutron excess that the 
selfconsistent mean field container is full up to the edge. In the 
inner crust of 
neutron stars where the nuclei form a Coulomb crystal these extra neutrons  
overflow into the interstitial space and form there a more or less 
dense neutron gas which also can be superfluid. In the inner crust the nuclei 
can actually turn into sheets and the neutron gas can form in between the 
sheets (a so-called lasagne configuration \cite{hae07}). As a first example 
we will 
treat in a schematic model such a slab configuration as is shown in 
Fig. \ref{slab1}, mostly because the 
quantal solution is readily available and, therefore, can serve as a test 
case for the validity of the TF approximation for treating the pairing 
problem. 
Indeed, we will find that 
the TF approach reproduces the quantal solution of the pairing properties, 
besides some shell fluctuations, very accurately. On the physical side, we 
point out that at the overflow point pairing can be strongly suppressed. This 
finding will then also be reproduced with a system where cold atoms are 
filled into a spherical container consisting of a narrow part at low filling, 
suddenly going over into a much wider container at higher chemical 
potentials, as it is displayed in the left panel of Fig. \ref{ketterle}
\cite{kett98} . 
A slightly different situation occurs with a double 
well potential, as the one shown in the left panel of Fig. \ref{bruno} . 
Again this potential is used in a slab configuration 
and TF and quantal results for the gaps are compared. \\

At the end, we return to nuclei in the crust of neutron stars where they are 
embedded in a more or less dense gas of neutrons. A Wigner-Seitz cell 
approximation will be applied to investigate this situation. Again similar 
features as in the previous examples will be found around the overflow point
in the transition from the outer to the inner crust.


\section{The formalism}

In this section, we will explain the TF approach to the pairing problem 
and apply it first to the slab configuration shown in Fig. \ref{slab1}.
Let us start out by writing down the usual BCS equations in three dimensions

\begin{equation}
\Delta_{\nu}=-\sum_{\nu'}V_{\nu,\nu'}\frac{\Delta_{\nu'}}{2\sqrt{(\varepsilon_{\nu'}
-\mu)^2 + \Delta_{\nu'}^2}},
\label{eq1}
\end{equation}

\noindent
where the $\varepsilon_{\nu}$'s are the single particle energies and $\mu$ the
chemical potential which can be used to fix the particle number
$N=\sum_{\nu}v_{\nu}^2$ with

\begin{equation}
v_{\nu}^2 = \frac{1}{2}\bigg[1 - \frac{\varepsilon_{\nu} - 
\mu}{\sqrt{(\varepsilon_{\nu}-
\mu)^2 + \Delta_{\nu}^2}} \bigg].
\label{eq2}
\end{equation}

The wave
functions and eigenenergies of a box as shown in Fig.~\ref{slab1} with a
potential-hole are given
in \cite{flu74} . For pairing, we use a contact force with a cut off $\Lambda$,
to make things simple. The single particle states in a slab configuration then
become $|\nu\rangle = |n, {\bf p}>$ where $n$ are the discrete quantum numbers
in transverse direction and ${\bf p}$ the momentum quantum numbers in slab
direction.
To obtain the gap equation in this case we start by integrating 
the gap equation (\ref{eq1}) over momenta 
in slab direction:

\begin{equation}
\Delta_n = -\sum_{n'} \int \frac{d^2p}{(2\pi \hbar)^2} V_{nn'} \Theta(\Lambda -
\varepsilon_{n'} - \varepsilon_p) \frac{\Delta_{n'}}{2E_{n'}(p)},
\label{eq3}
\end{equation}

\noindent
with $E_n(p) = \sqrt{(\varepsilon_n + \varepsilon_p - \mu)^2 + \Delta_n^2}$
the quasiparticle energy,
$\Theta(x)$ the step function, and $\varepsilon_n, \varepsilon_p$ being
the discrete
single particle energies in transverse direction and kinetic energies in
slab direction, respectively. After simple algebra, 
one arrives at the following gap equation for a slab configuration

\begin{equation}
\Delta_n = - \sum_{n'} \Theta(\Lambda - \varepsilon_{n'}) V_{nn'} K_{n'}.
\label{eq4}
\end{equation}
  \noindent
The pairing tensor in equation (4) is then given by
\begin{equation}
K_n =\frac{m}{4\pi\hbar^2}\Delta_{n} \ln{\frac{\Lambda - \mu +
\sqrt{(\Lambda - \mu)^2 +
\Delta_n^2}}
{\varepsilon_n - \mu + \sqrt{(\varepsilon_n - \mu)^2 + \Delta_n^2}}},
\label{eq5}
\end{equation}


\noindent
where $m$ is the particle mass and  the indices $n$ stand for the level
quantum numbers in the confining
potential of the left panel of Fig.~\ref{slab1}.
The matrix elements
$V_{nn'}= -g \int_{-L}^{+L}|\varphi_n(z)|^2 |\varphi_{n'}(z)|^2 dz$ of the
pairing contact force
$v_{pair}({\bf r} - {\bf r}')= -g\delta({\bf r} - {\bf r}')$ used in this case 
can be evaluated straightforwardly from the wave functions 
$\varphi_n(z)$ given in \cite{flu74} .

Before we show the results, let us explain our Thomas-Fermi (TF)
approach for this problem.
In the weak coupling regime, we have $\Delta/\mu << 1$. In this case the
canonical basis \cite{RS} can be replaced by the 
Hartree-Fock or mean-field one:

\begin{equation}
H|n \rangle = \epsilon_n|n \rangle.
\label{eq6}
\end{equation}

\noindent
At equilibrium and for time reversal invariant systems canonical conjugation
and time reversal operation are related by

\begin{equation}
\langle{\bf r}|{\bar n}\rangle =
\langle n|{\bf r}\rangle \Rightarrow
\langle{\bf r}_1 {\bf r}_2|n {\bar n}\rangle =
\langle{\bf r}_1|{\hat \rho_n}|{\bf r}_2\rangle,
\label{eq7}
\end{equation}

\noindent
with $\hat \rho_n = |n\rangle \langle n|$. For the pairing matrix element,
we, therefore, can write

\begin{equation}
V_{n n'} = \langle  n {\bar n}|v|n' {\bar n'} \rangle
=\int \langle{\bf r}_2|{\hat \rho_n}|{\bf r}_1 \rangle
\langle{\bf {r_1}} {\bf {r_2}} \vert v \vert {\bf {r_1'}} {\bf {r_2'}}\rangle
\langle{\bf r'}_1|{\hat \rho_n}|{\bf r'}_2 \rangle
{d\bf {r_1}}{d\bf {r_2}}{d\bf r_1'}{d\bf r_2'}.
\label{eq8}
\end{equation}

\noindent
The Schroedinger equation (\ref{eq6})
can be writen in terms of $\hat \rho_n$ as

\begin{equation}
 (H - \epsilon_n)\hat \rho_n = 0.
\label{eq9}
\end{equation}

\noindent
Taking the Wigner transform of this latter equation, we obtain in the
$\hbar \rightarrow 0$ limit the following c-number equation 
\cite{RS} :
$(H_{cl.} - \epsilon)f_{\epsilon}({\bf R}, {\bf p}) = 0$.
The solution of this equation in the sense of distribution theory is
with $x\delta(x)=0$ given by

\begin{equation}
f_E({\bf R},{\bf p}) = \frac{1}{g^{TF}(E)}\delta(E - H_{cl.}) + O(\hbar^2).
\label{eq10}
\end{equation}

\noindent
with

\begin{equation}
\nonumber
H_{cl.} = \frac{p^2}{2m^*({\bf R})} + V({\bf R}) \quad \mbox{and} \quad
g^{TF}(E) = \frac{1}{(2\pi \hbar)^3} \int  d {\bf R} d {\bf p} \delta(E -
H_{cl.}).
\end{equation}

\noindent
with $m^{*}({\bf R})$ the effective mass and $V({\bf R})$ the mean field 
potential.
Equation (\ref{eq10}) means that the phase space 
distribution corresponding to a state
$|n\rangle$ at high energy is concentrated around the classical
energy shell that, indeed, is a well known fact.

\noindent
The TF version of the gap equation (\ref{eq4}) then reads

\begin{equation}
\Delta(E) = - \int_{V_0}^{\Lambda} dE' g(E') V(E,E') K(E')
\label{eq11}
\end{equation}
with $K(E)$ an obvious generalisation of $K_{n}$ in (\ref{eq5}).
The matrix elements $V(E,E')$ can be evaluated in replacing $|\varphi_n(z)|^2$
by \cite{vin03}

\begin{equation}
\rho^{TF}_E(z) = \int \frac{dp}{2\pi \hbar} f_E(z, p) =
\frac{1}{g^{TF}(E)}\frac{1}{2\pi}
\big(\frac{2m}{\hbar^2} \big)^{1/2}
[E - V(z)]^{-1/2},
\label{eq12}
\end{equation}

\noindent
which is
the on-shell TF density in transverse direction (please note that we are
in a 1D case here, contrary to what is treated above where it is 3D). As
the reader will easily realise,
the way of proceeding is very different from usual LDA where the finite size
dependence is put into the (local) chemical potential, $\mu(z) = \mu - V(z)$,
whereas here it is put into the matrix elements of the pairing force
(notice that in LDA they are computed using plane wave functions).

\begin{figure}
\includegraphics[height=5.5cm,angle=-90]{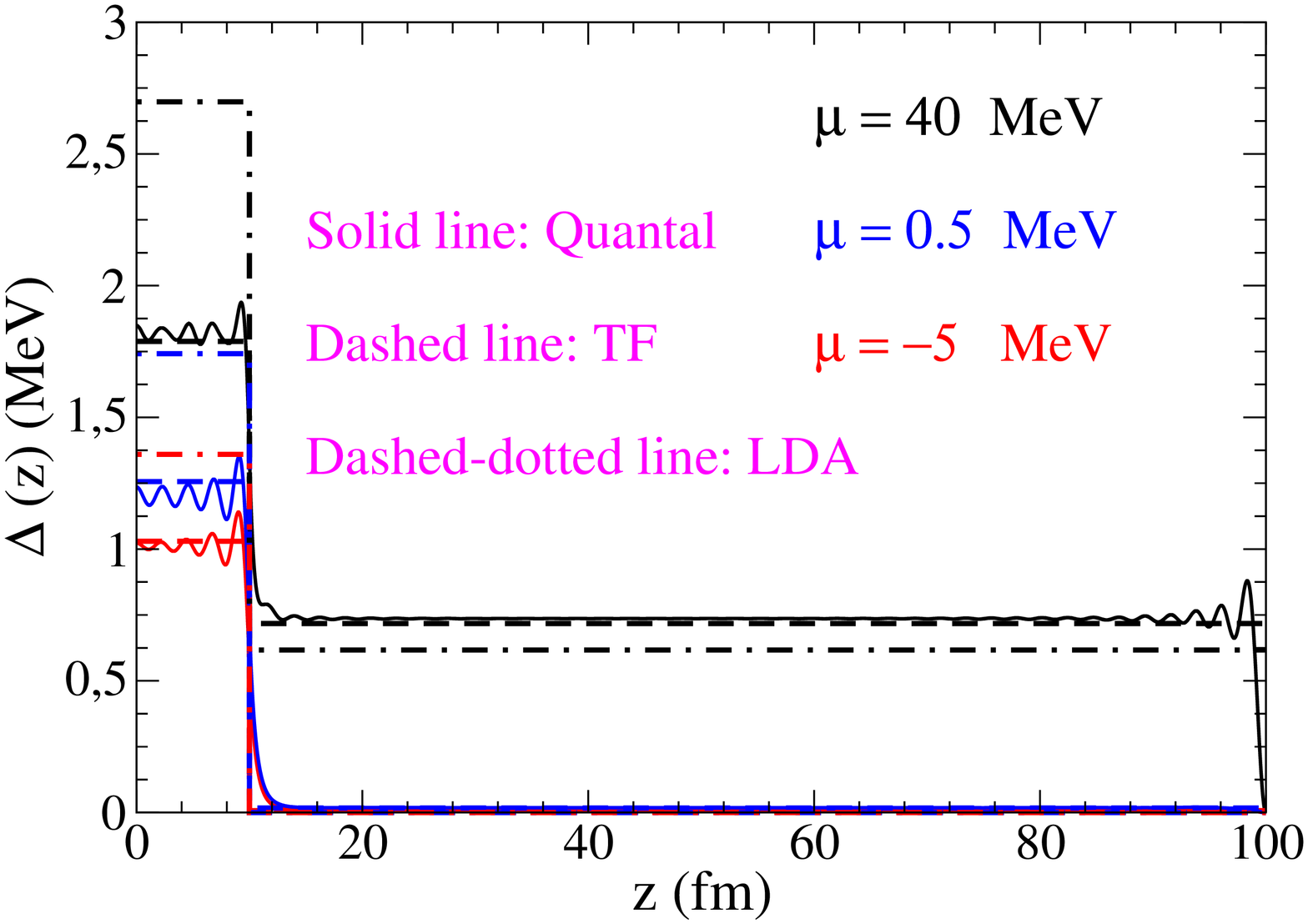}
\includegraphics[height=5.5cm,angle=-90]{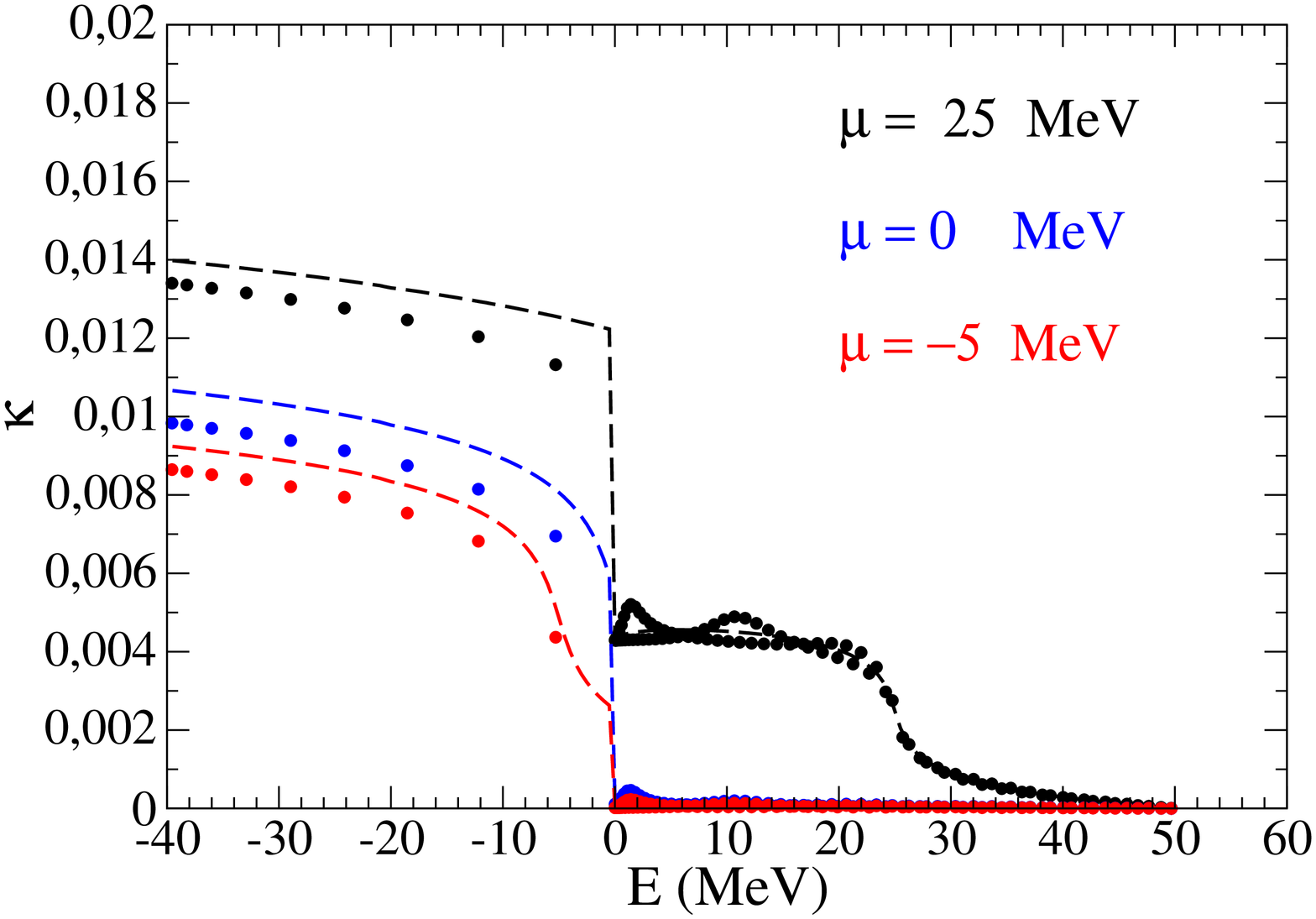}
\caption{(Coloronline)
\label{slab3}
Left panel: Position dependence of the gap in the slab geometry
for different values of the chemical potential. Quantal, TF, and LDA results
are shown. Notice that $\Delta$ for
$\mu$ = 0.5 and -5.0 MeV is practically zero in the gas region.
Right panel: Comparison of quantal (dots) and TF (broken lines) values of the
pairing tensor $K$. }
\end{figure}

\begin{figure}
\includegraphics[height=5.5cm,angle=-90]{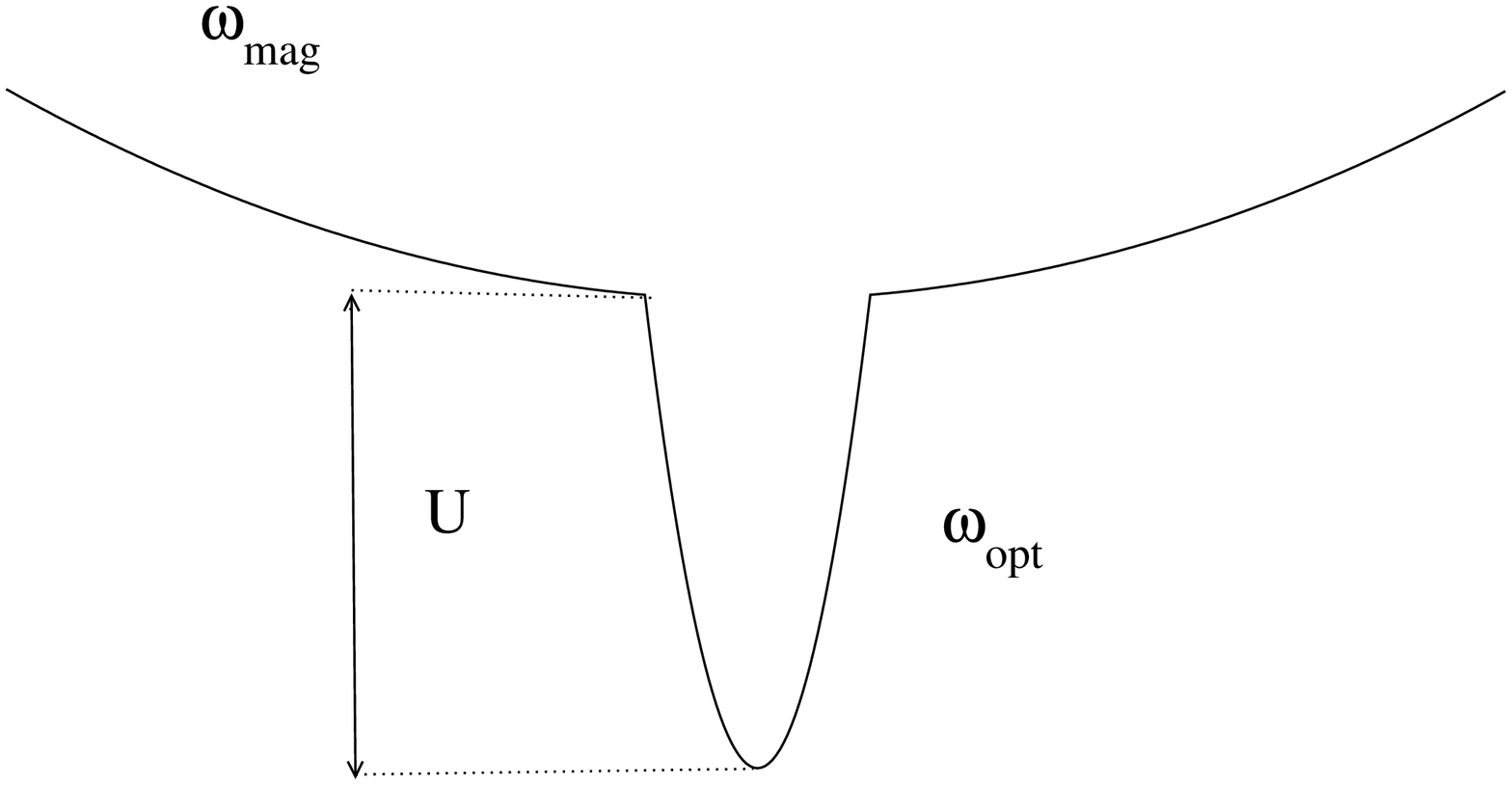}
\includegraphics[height=5.5cm,angle=-90]{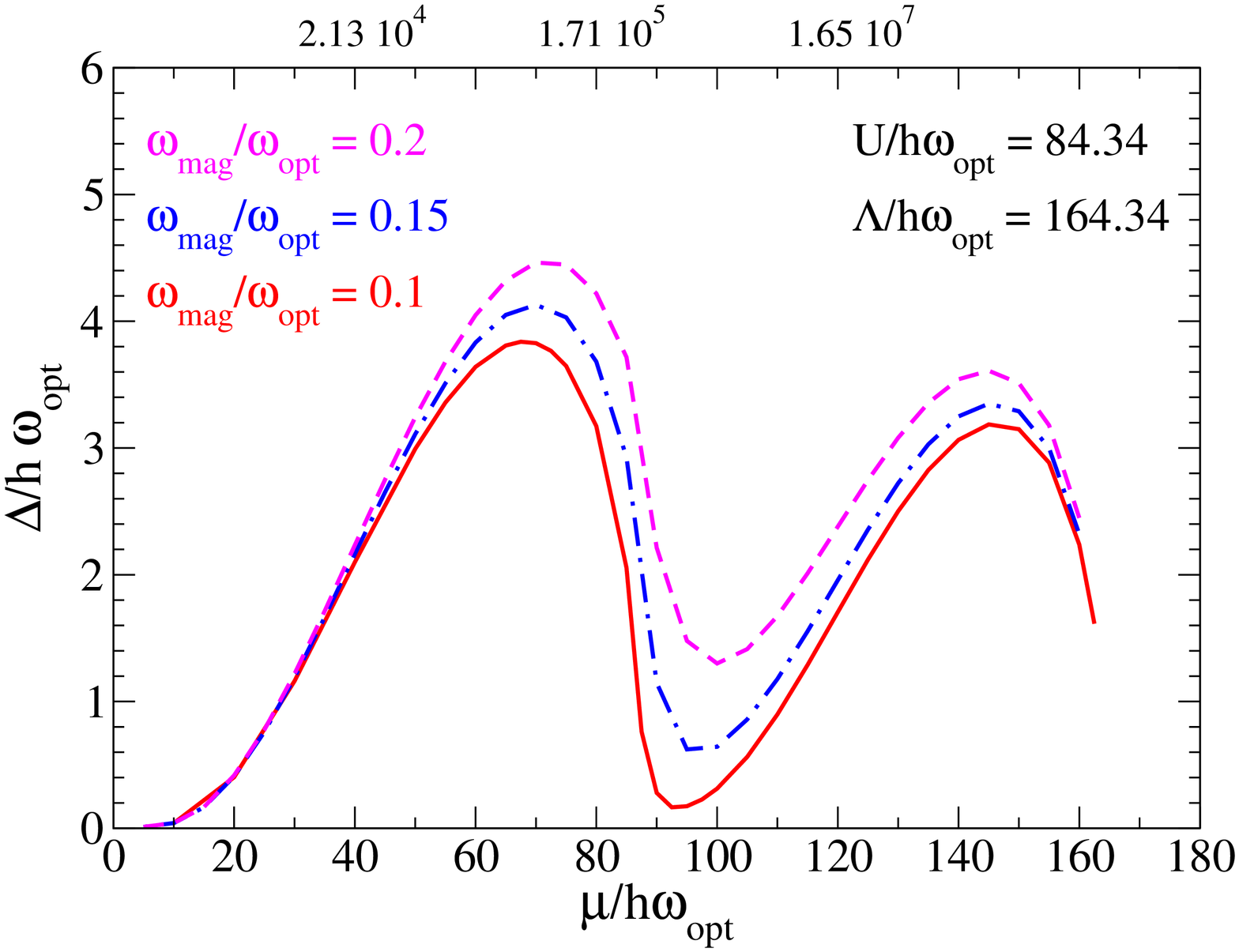}
\caption{(Coloronline)
\label{ketterle}
Average TF gaps at the Fermi energy as a function of
the chemical potential (right panel) for the potential shown in the left panel.
In the completely filled
optical trap ($\mu=U$) we accomodate 10$^5$ atoms in each spin state.
The total number of atoms in the trap with
$\mu/\hbar\omega_{opt}$=40, 80 and 120 are indicated in the upper
horizontal axis.}
\end{figure}

\begin{figure}
\begin{center}
\includegraphics[height=5.5cm,angle=-90]{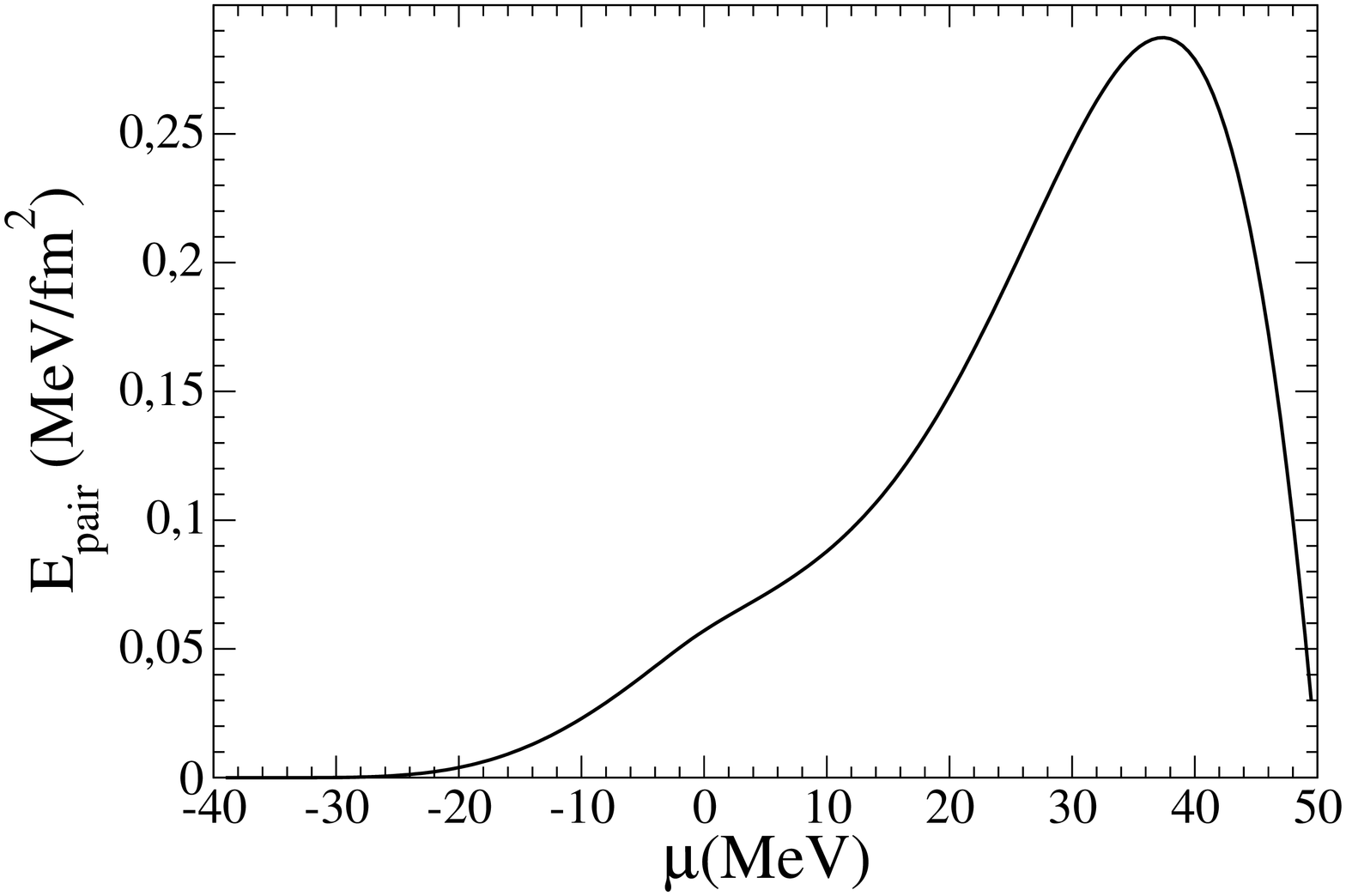}
\includegraphics[height=5.5cm,angle=-90]{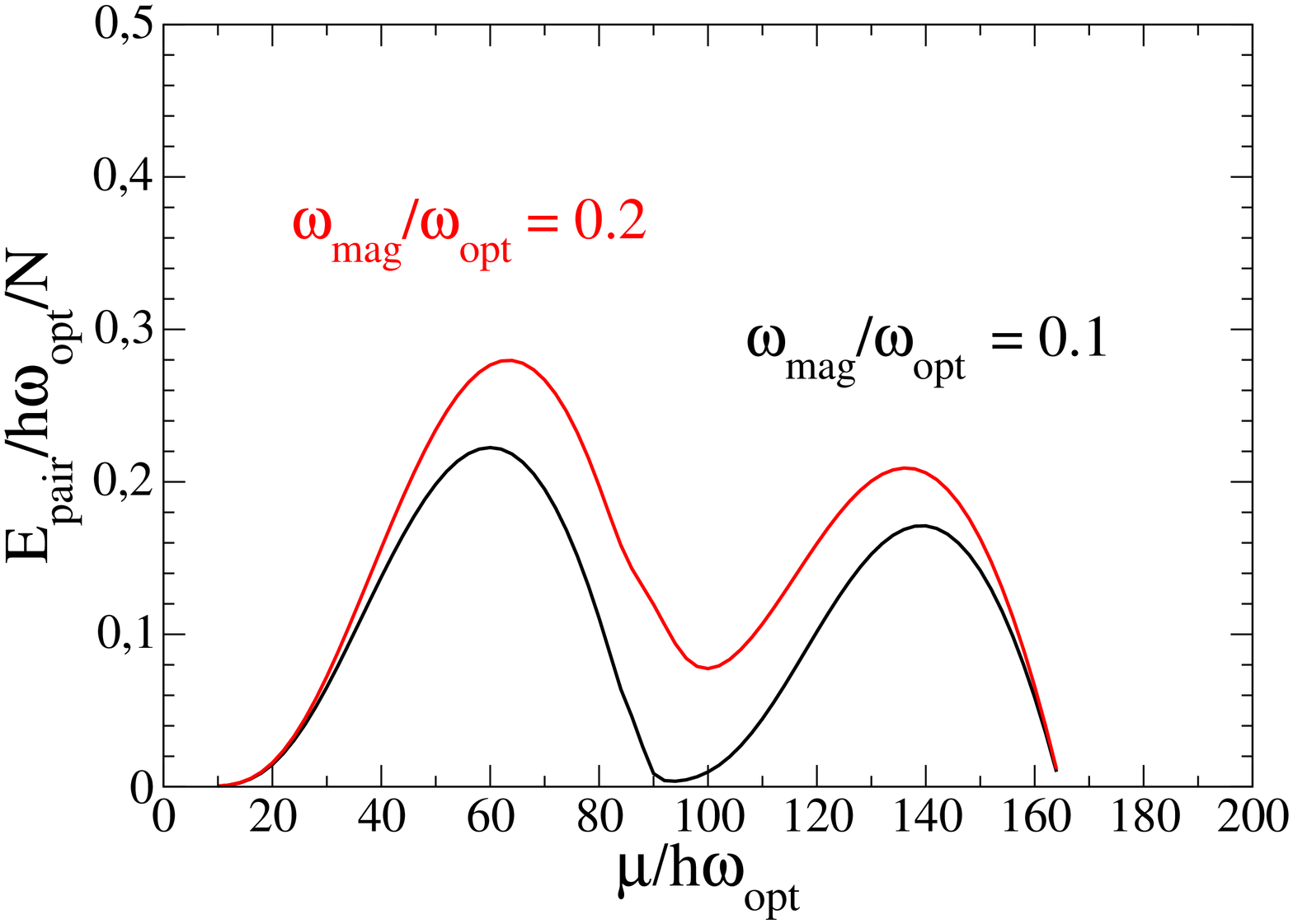}
\caption{(Coloronline) \label{epair} Pairing energy as a function of the 
chemical potential for the slab (left panel) and the double harmonic potential
discussed in the text (right panel).}
\end{center}
\end{figure}

\begin{figure}
\begin{center}
\includegraphics[height=5.5cm,angle=-90]{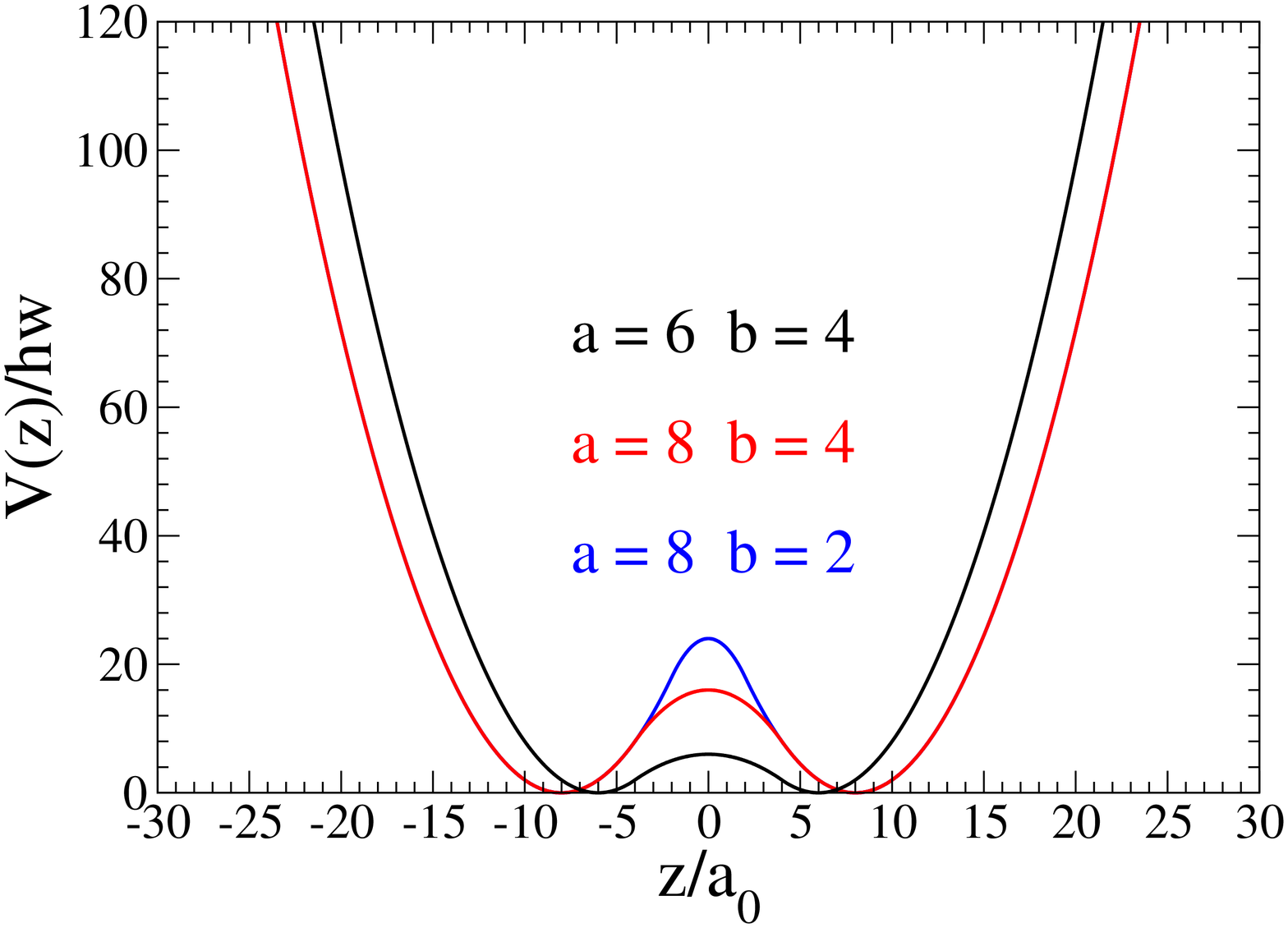}
\includegraphics[height=5.5cm,angle=-90]{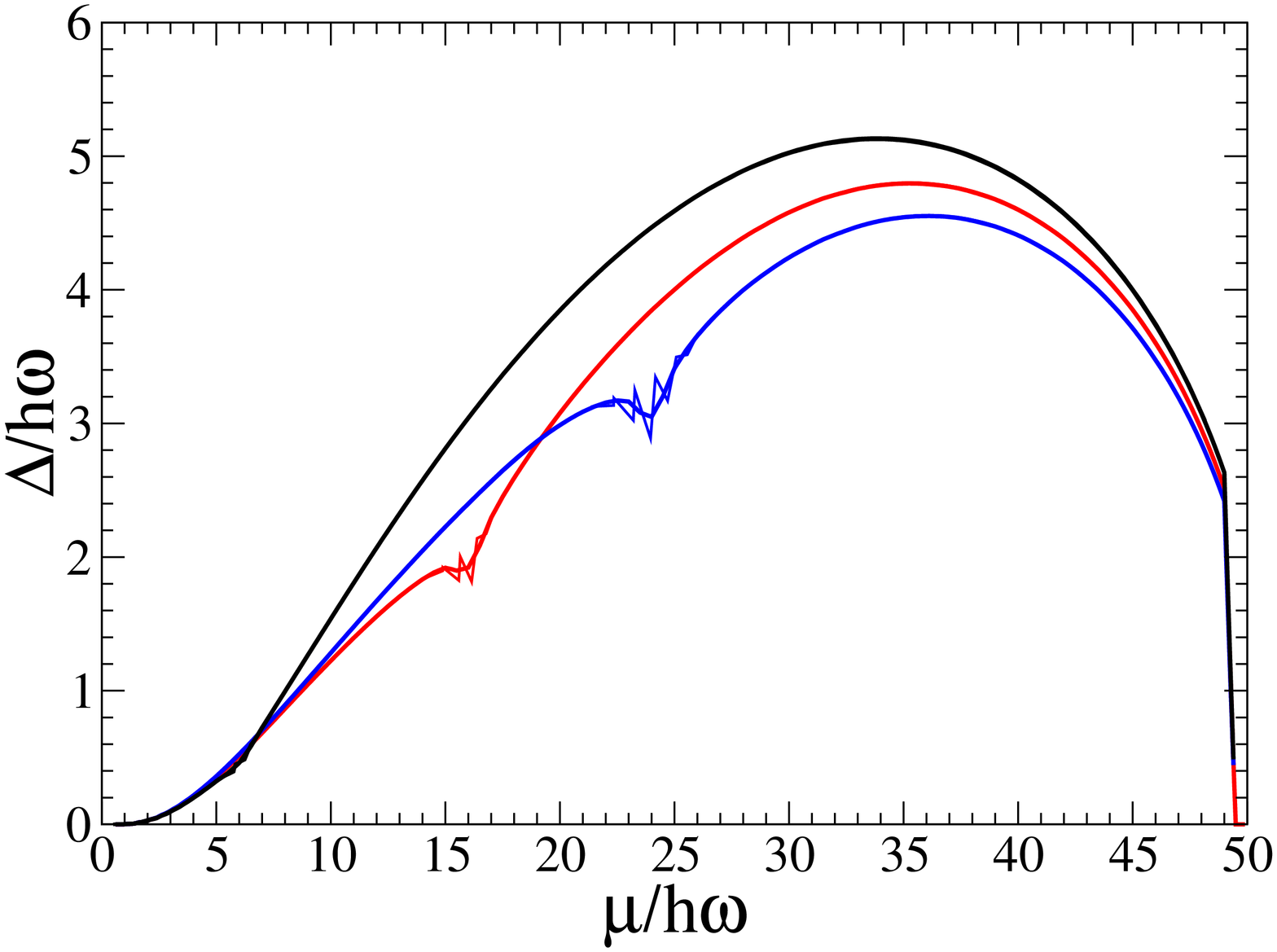}
\end{center}
\caption{(Coloronline) \label{bruno} Double well potential in slab geometry (left)
and gaps as a function of the chemical potential (right).
The different curves correspond to the parameters a=6 and b=4, a=8 and
b=4 and a=8 and b=2 (see text for explanation). Please note that TF and 
quantal values cannot be distiguished on the scale of the figure, besides 
in the region of the dips where TF passes through the average of the quantal 
oscillations.}
\end{figure}


\begin{figure} 
\begin{center}
\includegraphics[height=5.5cm,angle=-90]{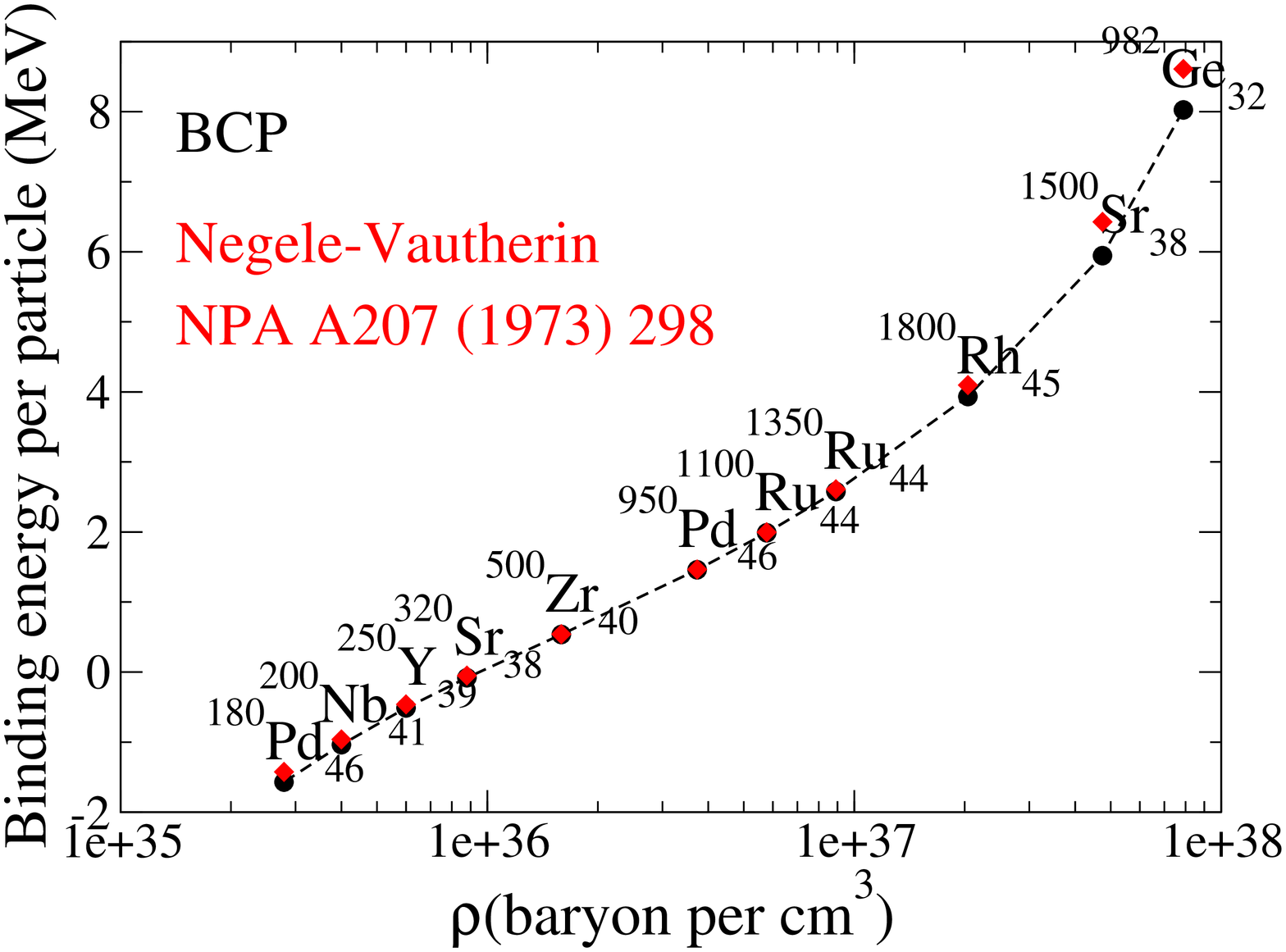}
\includegraphics[height=5.5cm,angle=-90]{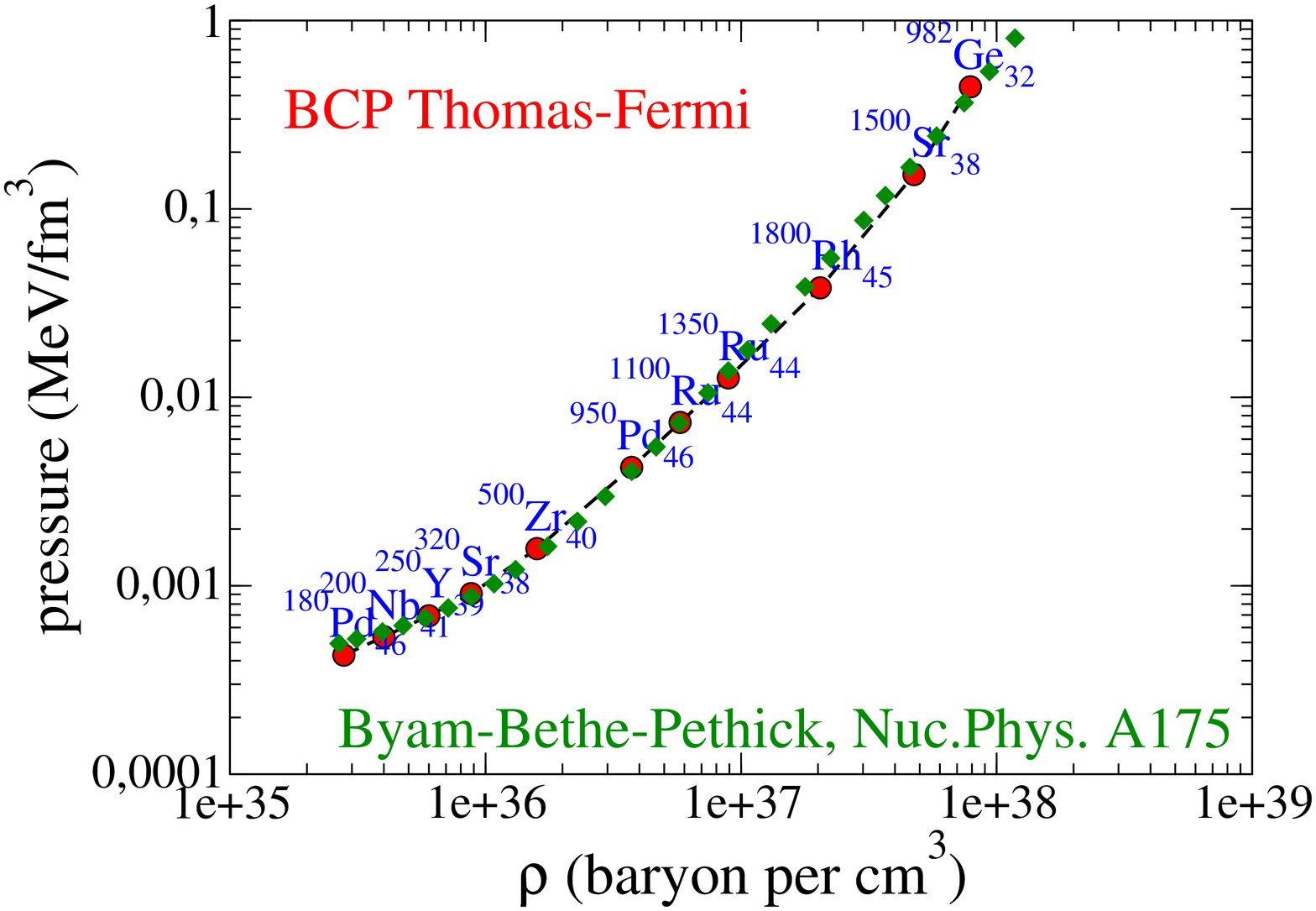}
\caption{(Coloronline) \label{bcpeos} Left: Binding energies per particle as a
function of average density in a Wigner-Seitz cell.
Red dots indicate quantal Skyrme HF calculations by Negele-Vautherin and
black dots correspond to semiclassical results with the variational
Wigner-Kirkwood (VWK) method and the Gogny D1S force. Right: BCP equation of 
state for the inner crust compared with the predictions of the 
Baym-Bethe-Pethick one.} \end{center}
\end{figure}

\begin{figure}
\includegraphics[height=5.5cm,angle=-90]{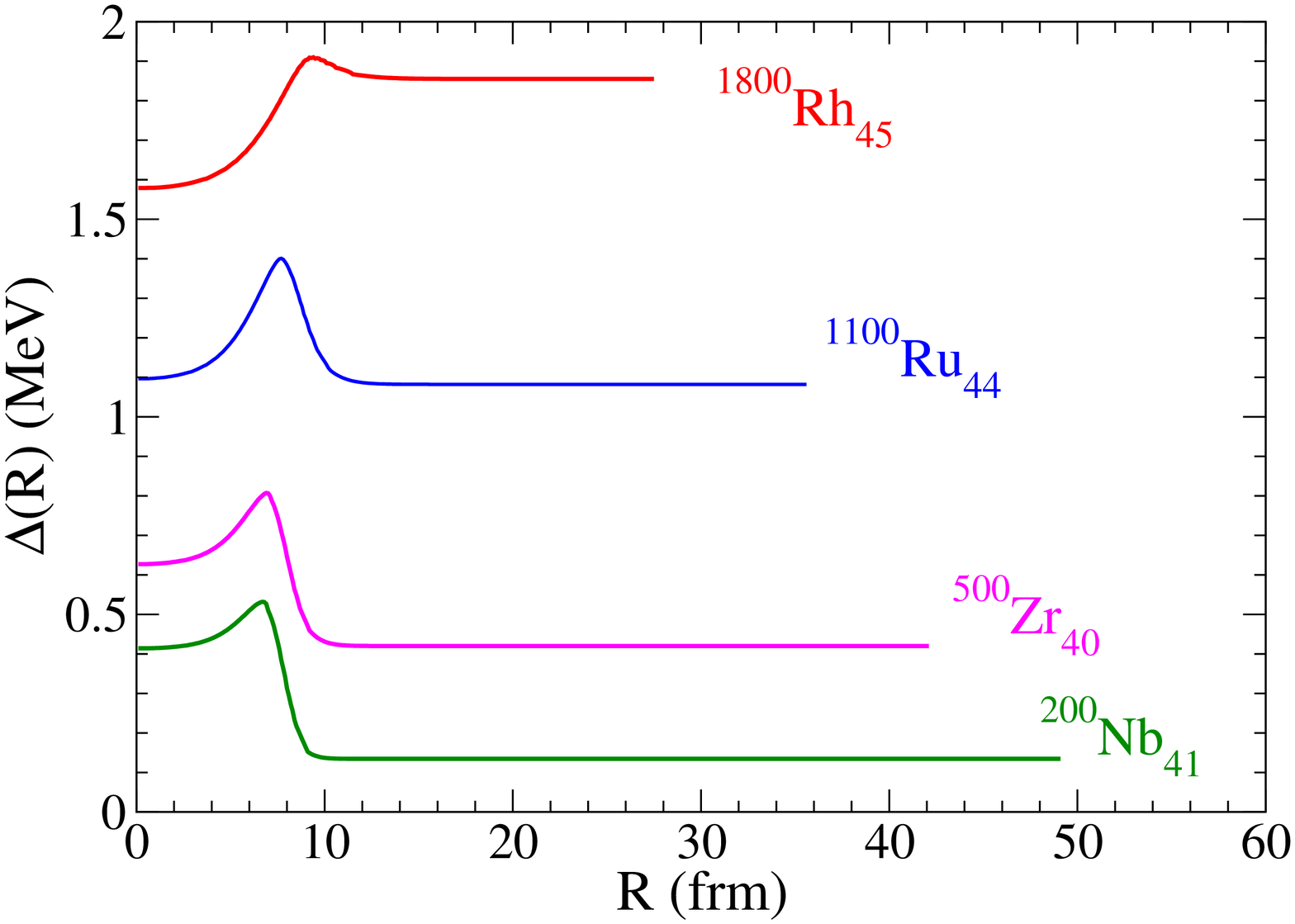}
\includegraphics[height=5.5cm,angle=-90]{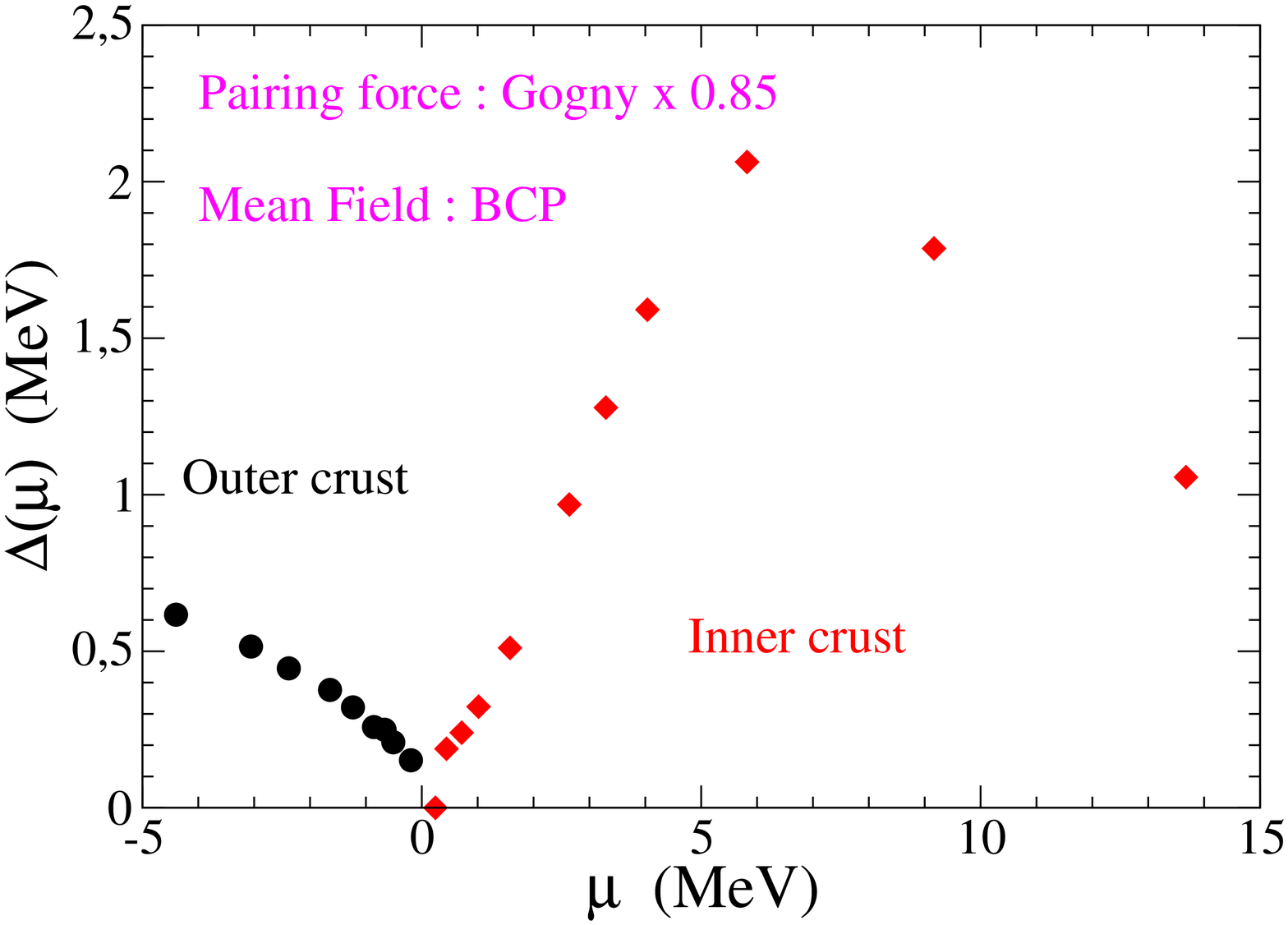}
\caption{(Coloronline) \label{gapcrust1} Left panel: Radial dependence
of the TF gaps in the
considered WS cells. The end points indicate the radius of the WS cells.
Right panel: 
Gap values for the Wigner-Seitz cells
corresponding to Fig.~\ref{bcpeos}. The zero point indicates the transition from isolated
nuclei (black dots) to the dripped case (red diamonds).}
\end{figure}

\begin{figure}
\includegraphics[height=5.5cm,angle=-90]{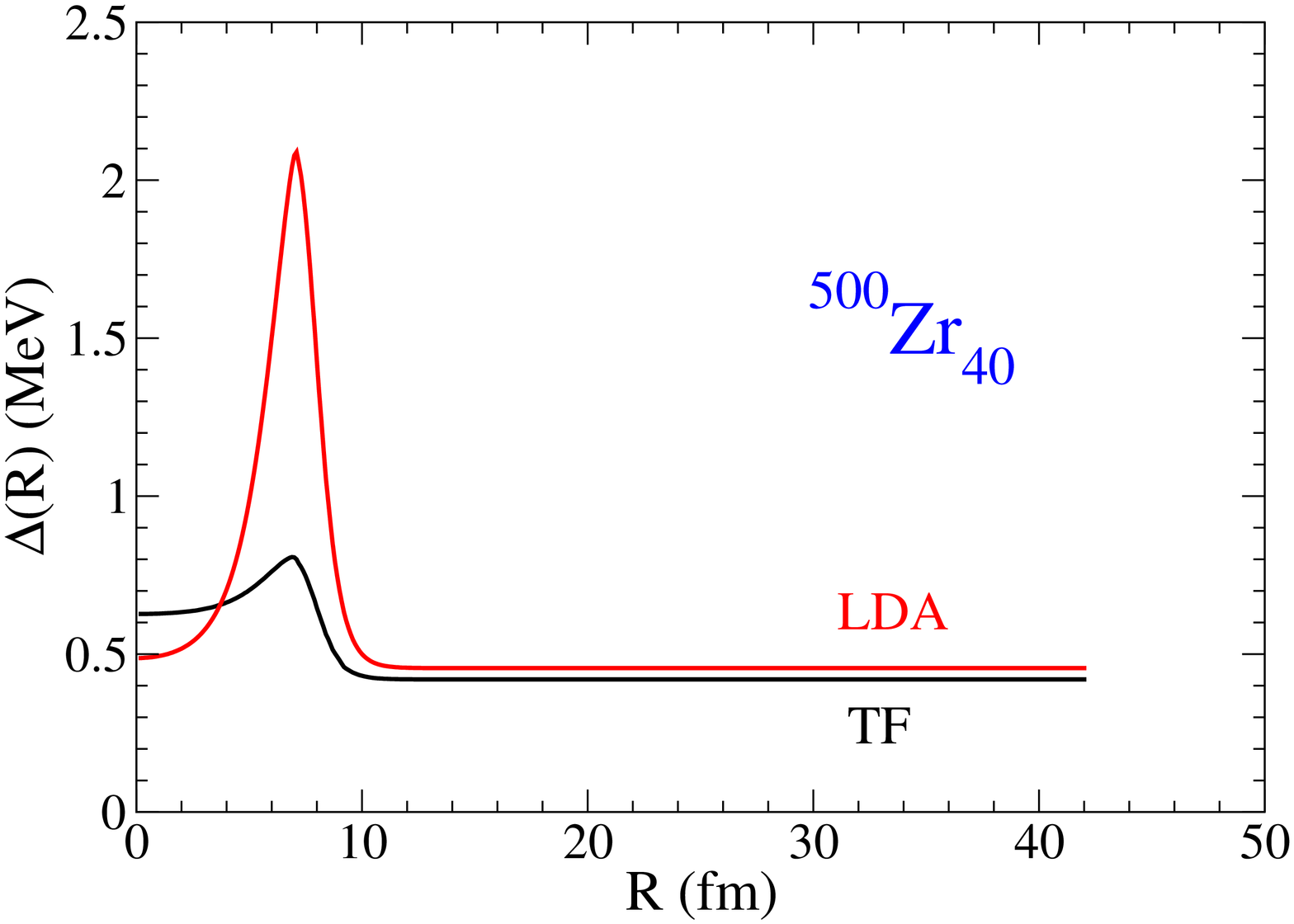}
\includegraphics[height=5.5cm,angle=-90]{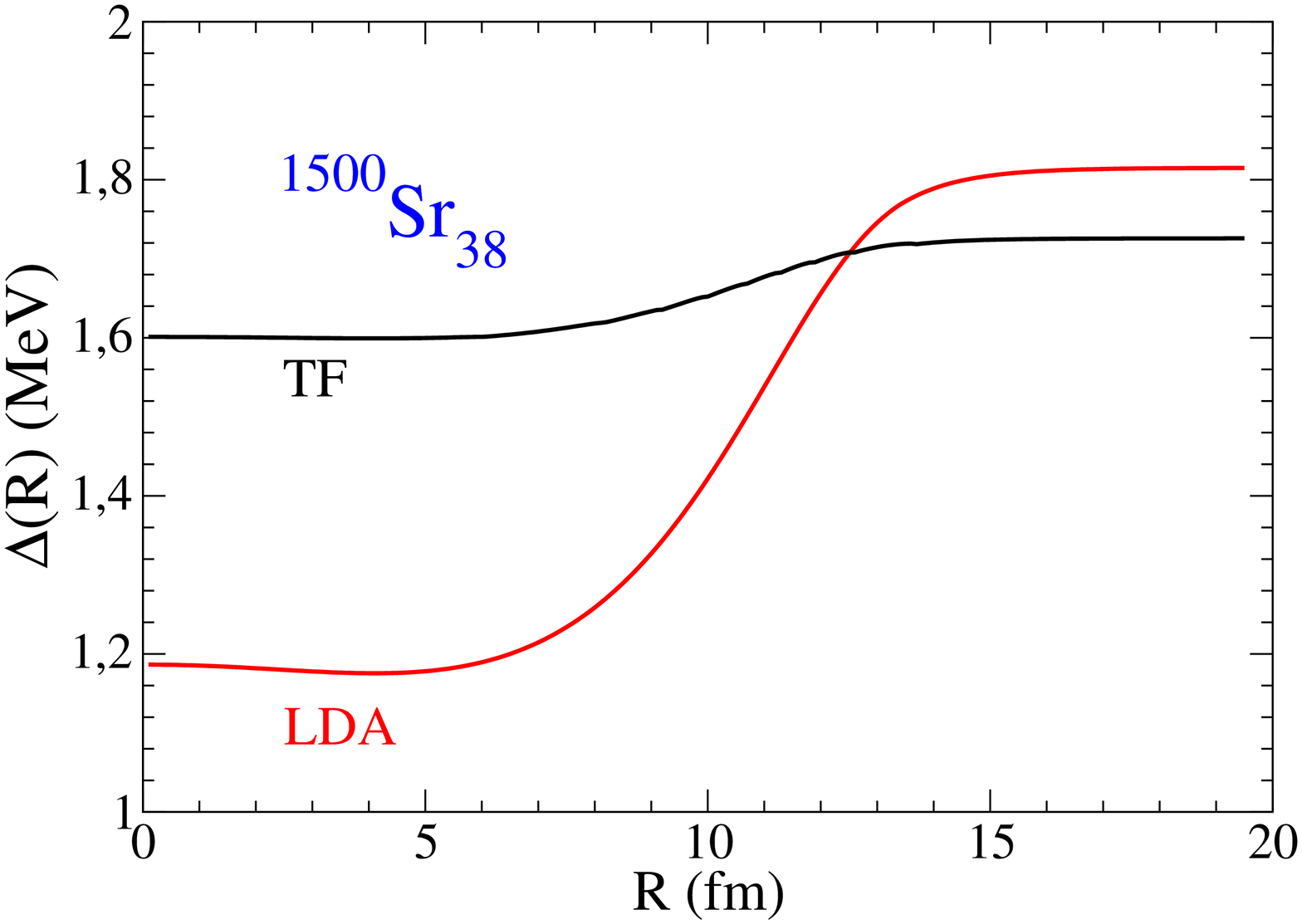}
\caption{(Coloronline) \label{gapcrust2} Comparison between TF and LDA 
gaps as a function of the position in a WS cell containing a single 
$^{500}_{40}$Zr nucleus (left) and $^{1500}_{38}$Sr (right).}
\end{figure}

\begin{figure}\begin{center}
\includegraphics[height=5.5cm,angle=-90]{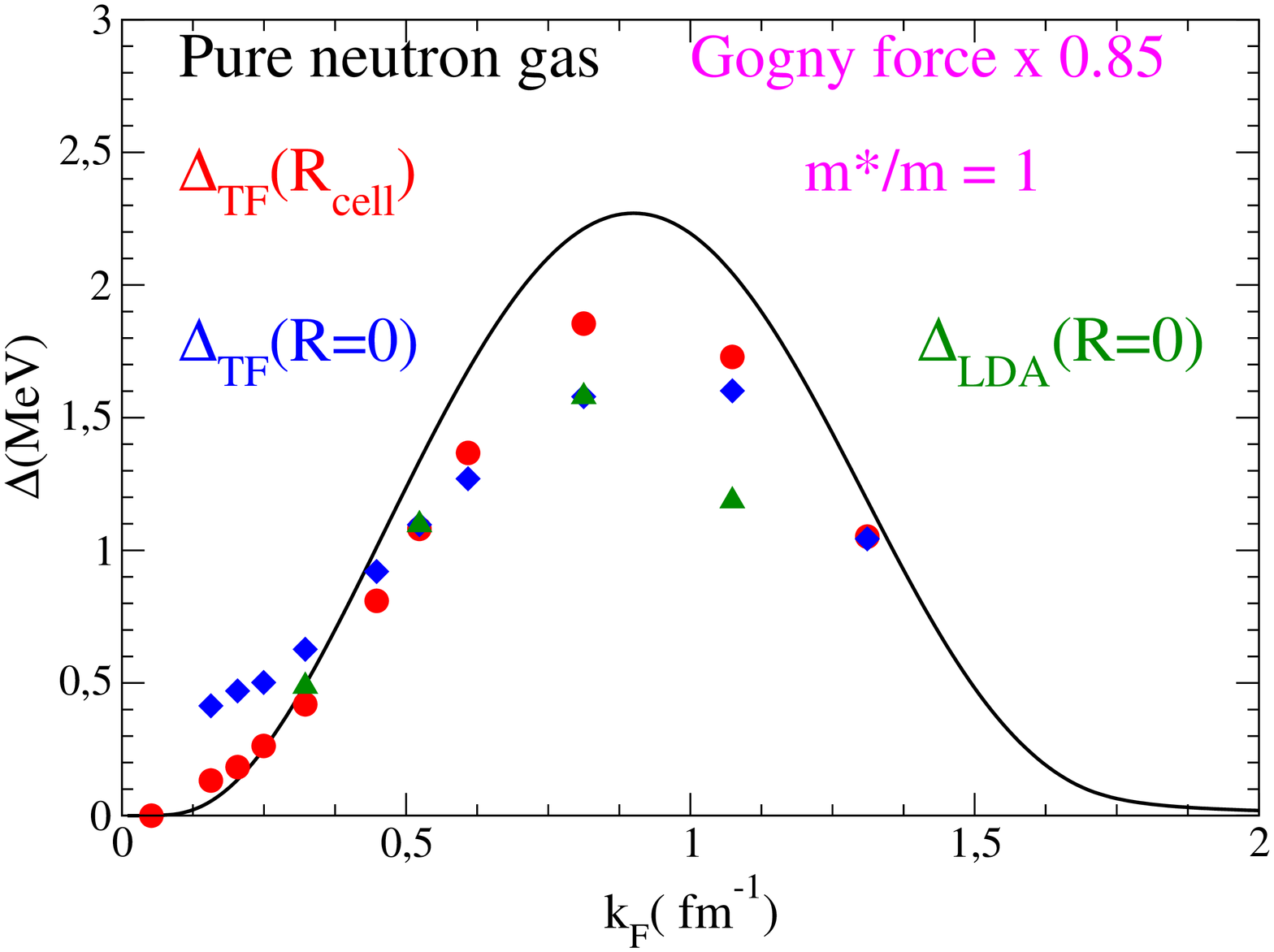}
\includegraphics[height=5.5cm,angle=-90]{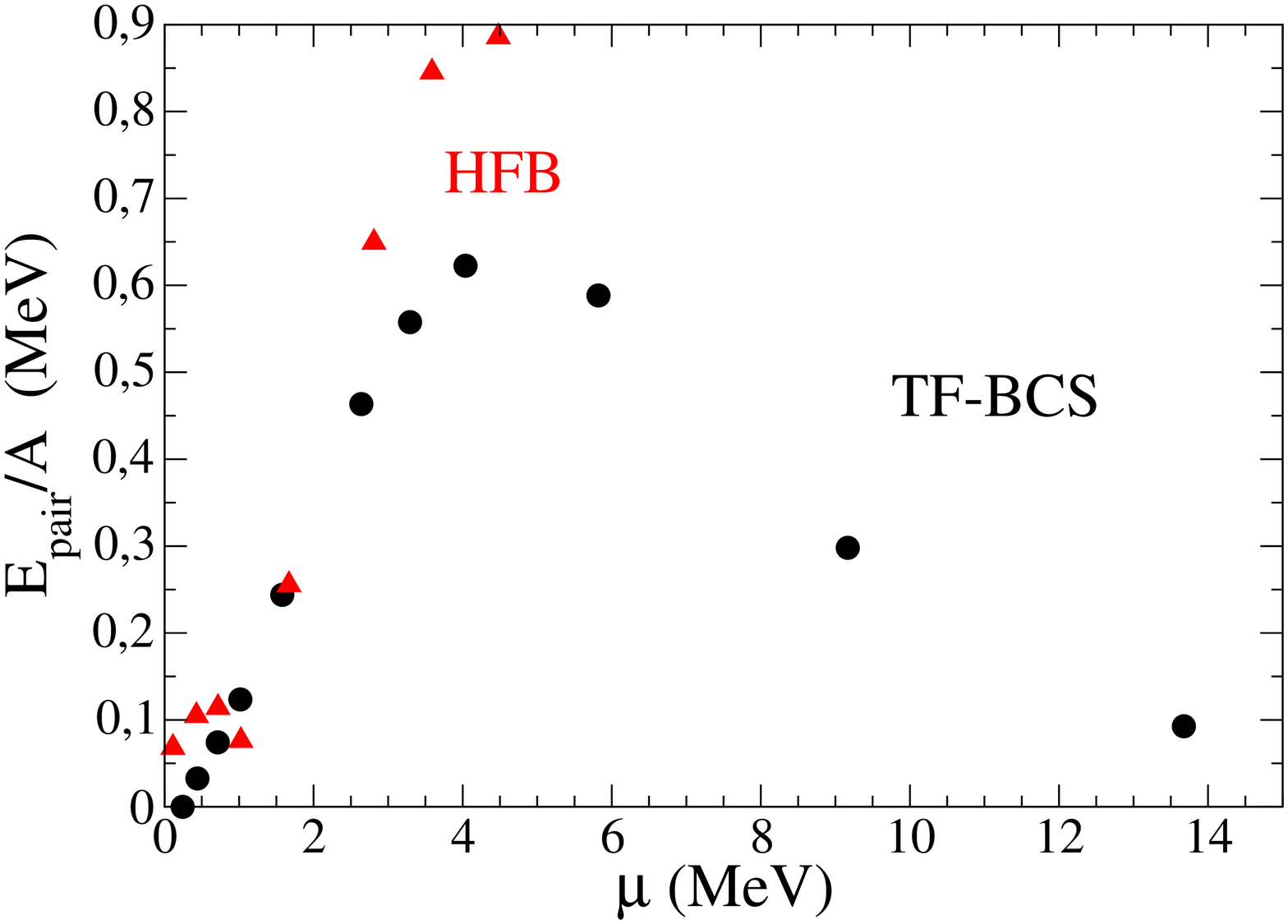}
\caption{(Coloronline) \label{gapcrust3} 
Left panel: The TF (blue diamonds) and LDA (green triangles) gaps at the
origin are compared with the gaps of the free neutron gas ($\Delta(R_{cell})$,
red dots) as a function of the Fermi momentum of the neutron gas 
at the edge of the cell.
Right panel: pairing energy per nucleon in
the inner crust. The TF results are compared with the quantal HFB values of 
ref.\cite{grill11a}}.
\end{center}
\end{figure}


\begin{figure} 
\includegraphics[height=5.5cm,angle=-90]{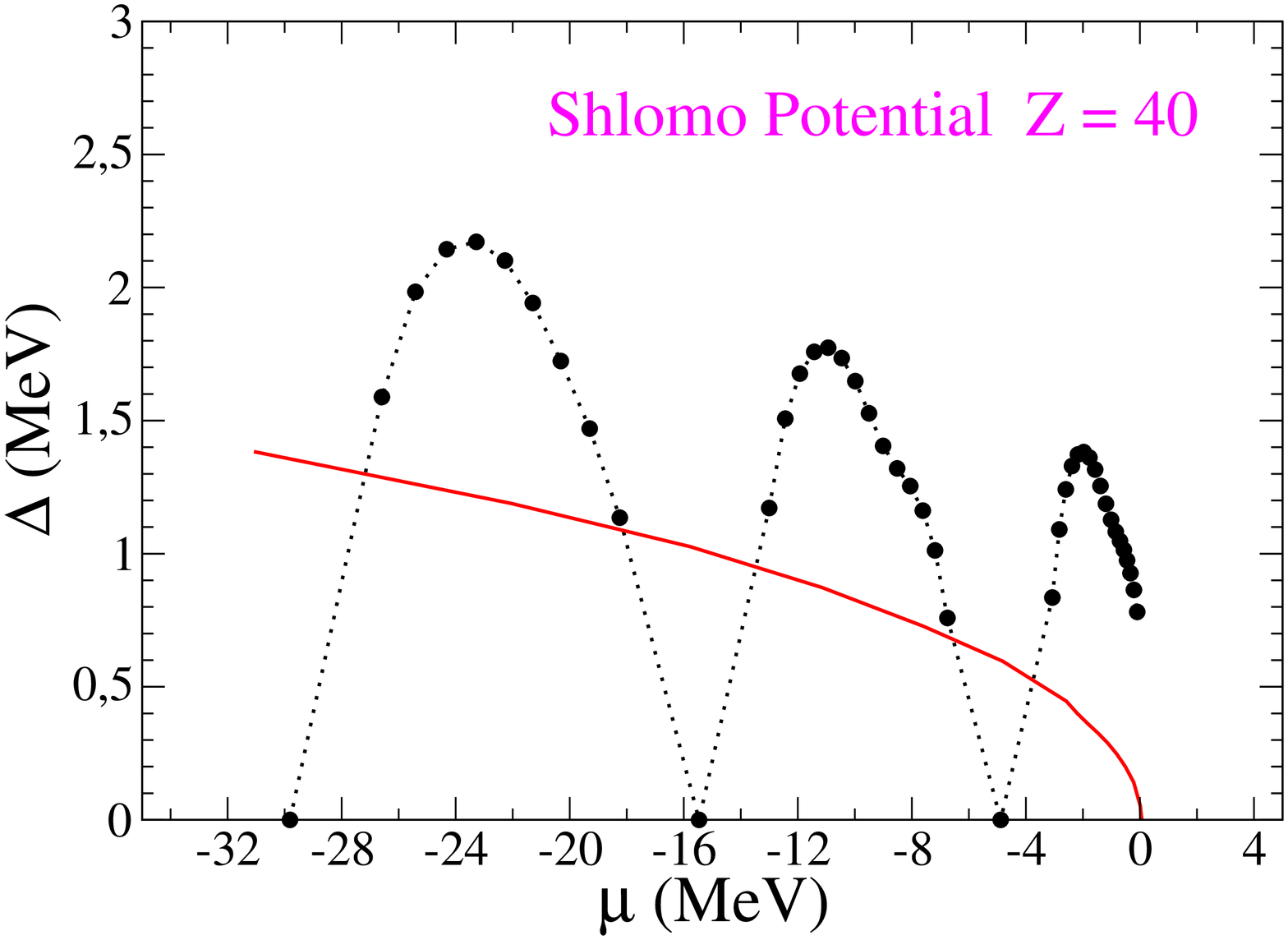}
\includegraphics[height=5.5cm,angle=-90]{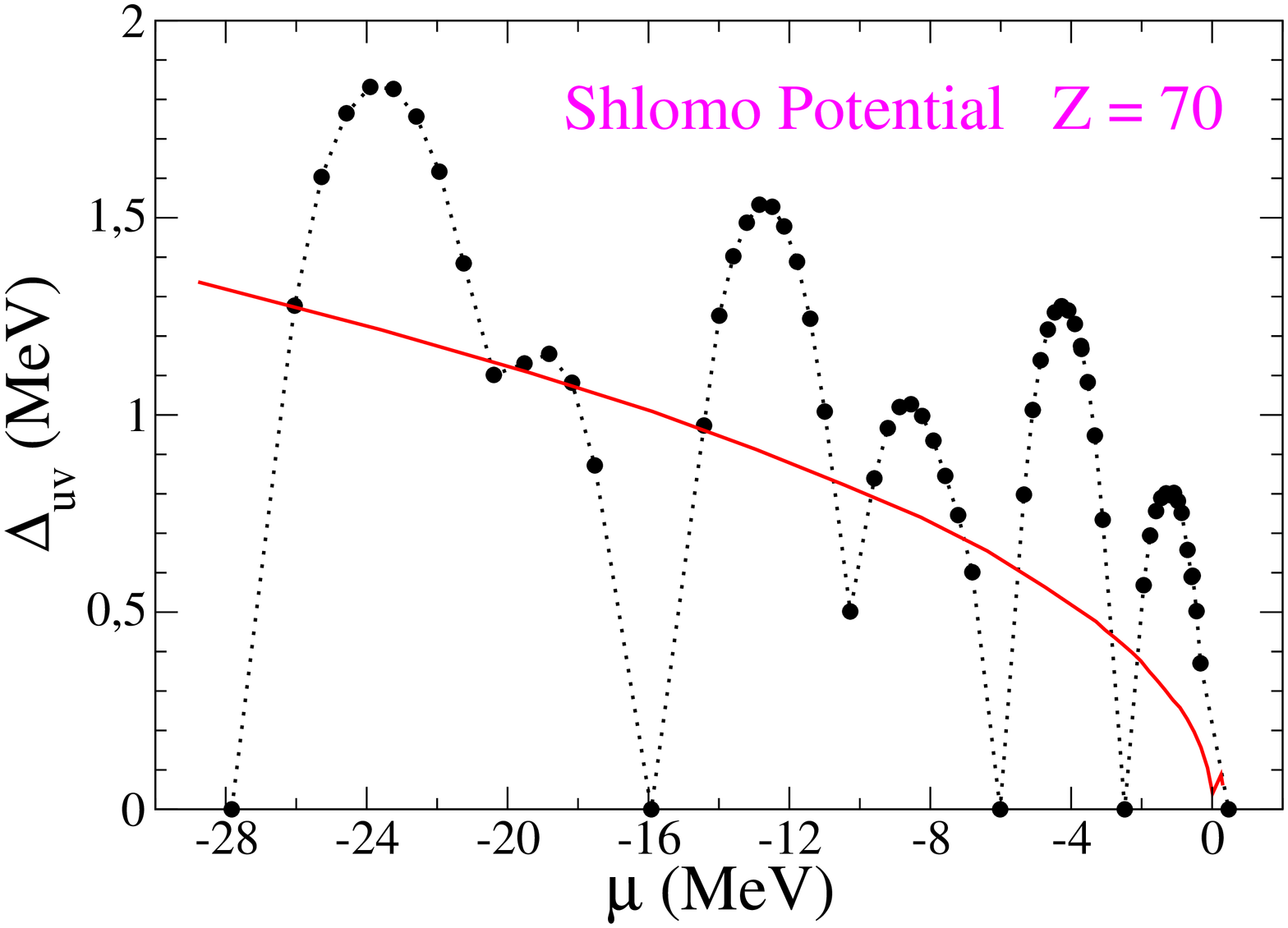}
\caption{(Coloronline) \label{Shlomogap}
Neutron  pairing gaps averaged with the pairing tensor ($uv$) 
along the $Z$=40 and $Z$=70 isotopic chains 
obtained with the Shlomo potential as mean field and the finite range
Gogny force renormalized by an attentuation factor of 0.85 in 
the pairing channel. Filled dots 
correspond to the BCS calculation and solid thick lines to the TF approach.}
\end{figure}

\begin{figure}
\includegraphics[width=4.25cm,angle=-90]{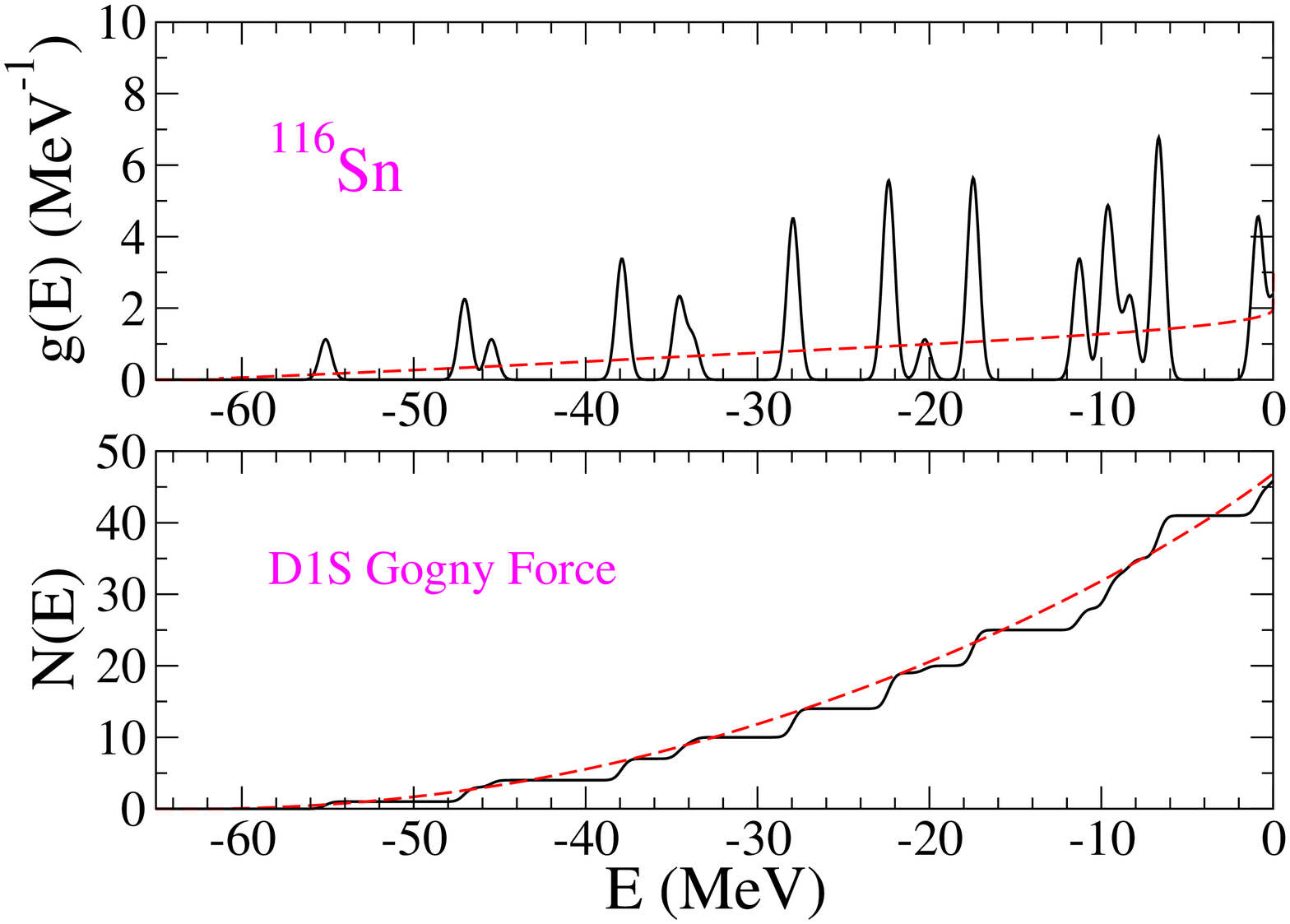}
\includegraphics[height=5.5cm,angle=-90]{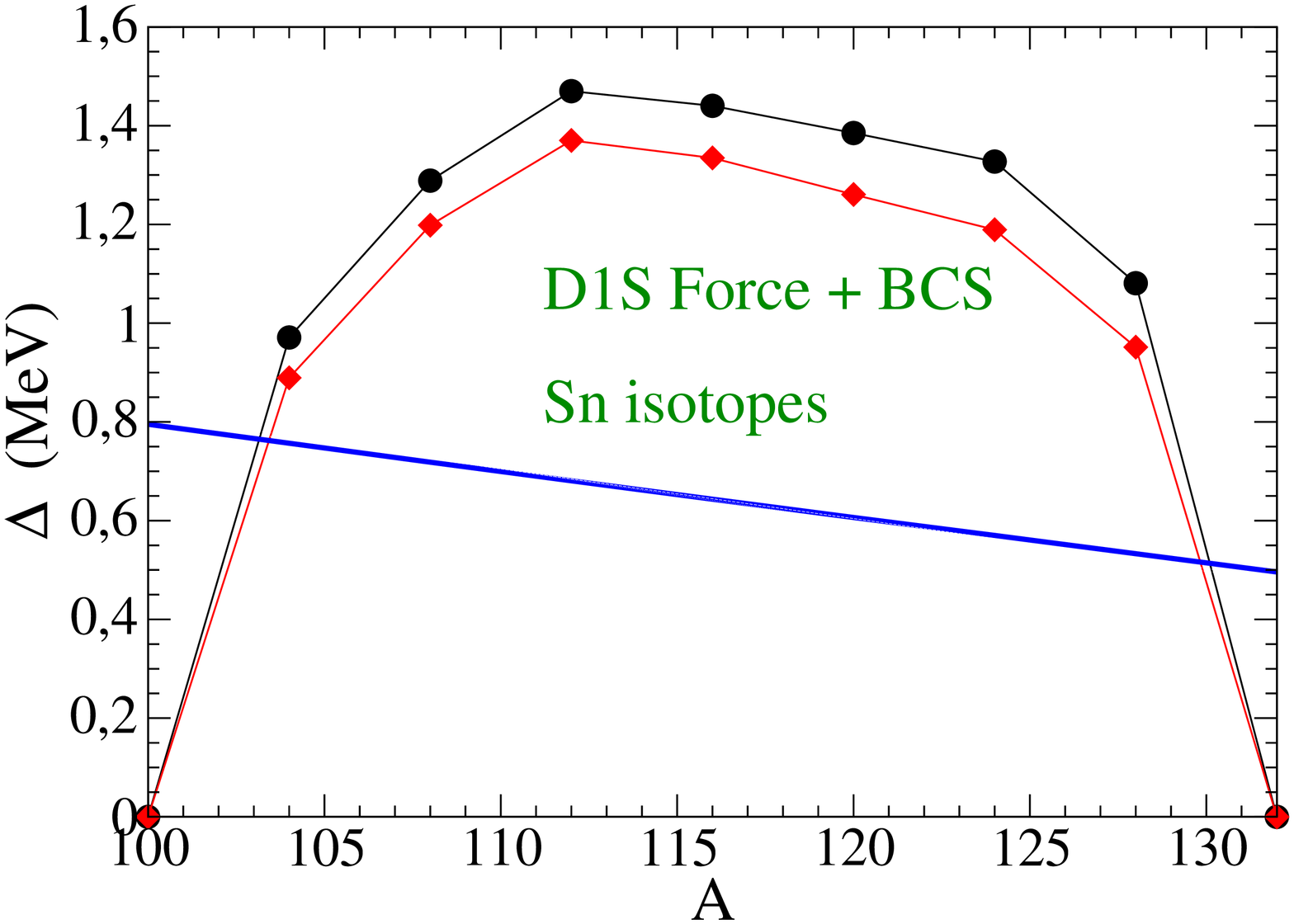}
\caption{(Coloronline) \label{D1Sgap} 
Left: TF (dashed line) and fluctuating (solid line) level density (upper panel)
and accumulated level density (lower panel). See text for details.
Right: Averaged gaps along the Sn isotopic chain computed fully quantally
at BCS level using the Gogny D1S force (circles), semiclassically using
the fluctuating level density (diamonds), and with the TF approach (solid line).}
\end{figure}

\section{Results}
We are now in a position to solve the quantal and TF gap equations
in the slab geometry for the confining potential displayed 
in the left panel of Fig.~\ref{slab1}. 
As an example we take as cut off $\Lambda$ = 50 MeV counted from the edge
of the pocket potential whose depth be $V_0=-$40 MeV.
Its extension ranges from  $-R$ to +$R$ with $R$ = 10 fm. The wide potential
with infinitly high walls has extension from $-L$ to +$L$ with $L$ = 100 fm.
For the coupling strength we take $g$= 150 MeV fm$^3$.
The result for the gap at the
chemical potential $\mu$ is shown in  the right panel of Fig. \ref{slab1} 
as a function of $\mu$. We start
with $\mu$ from the bottom of the pocket well, i.e. with zero density.
We then increase $\mu$, i.e. the density. We see that once the fill up of the
pocket reaches its top, the values of the gap sharply drop and practically
reach
zero. In the continuum the gaps slowly rise again. We see that quantal and TF
values are in close agreement. The overshoot of the TF solution for negative
$\mu$ very likely is due to the smallness of the pocket which only can
accomodate nine bound levels. The bunches of resonances in the continuum of 
the quantal solution are interesting but we did not try to explain them in this 
work. Before we come to an explanation of the
drop of the gaps at overflow (drip), let us study
the gaps as a function of position in transverse direction.
Quantally the position dependent gap is defined as:
$\Delta(z) = -g K(z)$ with $K(z) = \sum K_n |\varphi_n(z)|^2$.
Semiclassically, the relation between the gap and the pairing tensor becomes: 

\begin{equation}
 \Delta(z) = -g \int_{V_0}^{\Lambda} dE g^{TF}(E)
K(E) \rho^{TF}_E(z).
\label{eq13}
\end{equation}

In the left panel of Fig.~\ref{slab3}, we show the gap profiles for three 
different values of the chemical potential:
$\mu$ = 40, 0.5, and - 5 MeV. We see that quantal and TF results agree,
up to shell fluctuations, very well. We also show the LDA results.
They can be locally as wrong as by 50 percent. For other choices of 
system parameters
the LDA error may even be worse. This stems from the fact that in TF (and,
of course, also quantally), there is coupling between inside and outside the
pocket, i.e. the Cooper pair wave function extends into both regions what
tends to equilibrate the values of the gaps. In LDA the contrast is much
too strong. The drop of the gaps when crossing the threshold can be explained
by the fact that the single particle states are strongly delocalised in the
outer container and, thus, their contribution to the pairing matrix element
$V_{n,n'}$ becomes very small. In the right panel of Fig.~\ref{slab3} we show
the quantal and TF pairing tensors, $K_n$ and $K(E)$ respectively, 
defined before. We emphasize again the close agreement between quantal and TF results.

Having gained faith into our TF approach, we now can explore other
geometries and other systems, which are
more difficult for quantal solutions. In the right panel of Fig. \ref{ketterle}
 we display the result
for the gap $\Delta$ in the spherical double harmonic oscillator potential shown in
the left panel of Fig. \ref{ketterle}. The latter may be realised with cold 
fermionic atoms to study the overflow situation.
A zero range pairing force with strength $g$=-1.0 and cut off 
$\Lambda$=164.34
(in the corresponding optical trap units with $\omega_{opt}=2\pi \times 1000$
Hz taken from \cite{viv01}) is used in this case.
We see that the result is qualitatively similar to the slab
case, though in
this spherical geometry the dip does not quite reach zero and also is shifted
slightly to an energy above the break point of the potential. Note that
this
depends strongly on the choice of the ratio $\omega_{mag}/\omega_{opt}$
as it can be seen in the figure. Also the gap starts to decrease
towards the minimum quite early. This is contrary to what happens in the slab
case, where the change is very abrupt. The reason probably lying in the
spherical symmetry of the considered system.  
It would be interesting to
see whether our prediction can be verified experimentally.\\

It also is interesting to study the pairing energy, defined in TF 
approximation as $E_{pair} = \frac{1}{N}\int dE g(E) \Delta(E) \kappa(E)$. 
The results of the pairing energy per particle for the  slab potential, 
Fig. \ref{slab1}, and in  the H.O. case  of 
Fig. \ref{ketterle}  are shown in Fig. \ref{epair}. We see that
$ E_{pair}$ behaves quite differently in the two cases. In the spherical 
example for cold atoms the depression at the overflow point is also seen 
in the pairing energy whereas the depression is completely washed out in 
the case of the slab. The reason for this qualitative and strong difference 
must come from the fact that in the slab case the drop of 
the gap as a function of the chemical potential
$\Delta(\mu)$ at 
overflow is extremely steep, almost vertical. Furthermore, in the 
pairing energy 
corresponding to the slab, the pairing tensor $\kappa(E)$ should actually 
be replaced by $K(E)$ corresponding to (\ref{eq5}). Being integrated 
over the momenta in slab direction 
it does not show any peak at $E=\mu$ as is the case in the spherical case.
Therefore, in the integral of $E_{pair}$ also gap values further away from 
the overflow point are picked up which are not small at all.\\

Another interesting geometry which can be considered is a potential with a 
barrier at the origin, i.e. a double well potential, see the left hand panel 
of Fig. \ref {bruno}. We treat this exemple again in slab geometry 
as in Fig. \ref{slab1}. It roughly may mock up an oblate and very elongated 
trap potential for cold atoms with a double well potential in transverse 
direction. The results for the gaps $\Delta(\mu)$ are shown in 
{ the right hand panel of Fig. \ref{bruno}}.
This potential, which depends on two parameters $a$ and $b$,  is defined 
as follows: 
$V(z) = \frac{1}{2}(z+a)^2$ if $z\le b$, 
$V(z) = \frac{a(a-b)}{2} + \frac{1}{2}(1 - \frac{a}{b})z^2$ when 
$\vert z \vert \le b$ and 
$V(z) = \frac{1}{2}(z-a)^2$ for $z\ge b$.
As before, TF and quantal results are in excellent agreement. 
When the chemical potential reaches the top of the barrier, there occurs 
again a  reduction of the gaps, since the wave functions at the 
Fermi surface 
suddenly get more extended above the barrier. This is the same effect as in 
the previous examples, although it is in this case less pronounced. Such a 
double well 
potential allows for the creation of a Josephson current if the population in 
left and right well are out of balance \cite{sme97}. Our TF approach may strongly 
facilitate the description of this phenomenon in the case of cold
fermionic atoms.\\

Let us now make a more realistic study of Wigner-Seitz (WS) cells
to simulate the inner crust of neutron stars. In this approach one considers 
a single nucleus of $N$ neutrons and $Z$ protons inside a spherical box of radius 
$R_{cell}$ as well as a uniform background of $Z$ electrons 
to preserve the charge neutrality of the cell \cite{hae07} . 
The mean-field, as explained in \cite{vin11b} , is computed selfconsistently 
in the TF approach using the BCP energy density functional \cite{bal08} .
In this semiclassical calculation we consider the same WS cells 
and mass numbers as in 
the old quantal calculation of Negele and Vautherin \cite{neg73} . However,
as far as shell corrections are not included, in our semiclassical 
calculation, we take
as representative nucleus in each cell, the beta-stable one computed 
\`a la TF along the corresponding isotopic chain. This is why the 
atomic numbers 
$Z$ of the representative nuclei differ from the ones
reported in \cite{neg73} while their mass numbers $A$ coincide. 

It must be pointed out that the total energy per baryon obtained with our
TF approach is in very good agreement with the quantal values 
reported in \cite{neg73} , as it is explicitly discussed
in Ref.\cite{vin11b} and again shown here in the left panel of 
Fig.~\ref{bcpeos}.
As an additional test of our TF mean-field calculation, 
in the {right} panel of Fig.~\ref{bcpeos}, we also display 
the EOS (i.e. pressure as a function of the WS average density) in the 
inner crust obtained in our semiclassical calculation 
compared with the results 
provided by the Baym-Bethe-Pethick EOS \cite{bay71} which is considered
a benchmark in large scale neutron star calculations. We find  
excellent agreement between both calculations.

The semiclassical description of the WS cells including pairing correlations
at TF level is obtained from this mean-field using the finite range part
of the Gogny D1S force \cite{D1S} in the pairing
channel \cite{D1Sa} .
In the left panel of Fig.~\ref{gapcrust1}, we display the radial dependences 
of the gaps in some selected WS cells. It is seen that when the gap
 is small outside the region of the nucleus, then the gap
also is small inside the nucleus. This stems from the very large
coherence length where one neutron of a Cooper pair can be in the huge
volume of the gas and the other inside the small volume of the
nucleus (proximity effect). In this way the gas
imprints its behavior for the gap also inside the nucleus. Such a conclusion
was also given in a quantal Hartree-Fock-Bogoliubov (HFB) calculation
by Grasso et al. in \cite{gra08} what shows
that the here employed BCS approximation
apparently yields very similar answers as a full HFB calculation for WS
cells \cite{bal07,bal07a,pas11} . 
More precisely, let us point out that 
the gaps in the region of the nuclei, corresponding to the inner crust and 
displayed in this figure, are 
strongly affected by the neutron gas. To illustrate this fact, we display in the
right panel of Fig.~\ref{gapcrust1} the values of $\Delta(R=0)$ (blue 
diamonds) and 
$\Delta(R=R_{cell})$ (red filled circles), compared with the gaps of the 
free neutron gas (continuous black line) at 
the density corresponding to the edge of the cell $\rho(R_{cell})$.
The semiclassical TF gaps $\Delta(R=R_{cell})$, as expected, closely
follow the free neutron gas values in agreement with HFB calculations
\cite{pas11} . 
As seen, the gap values at the origin, $\Delta(R=0)$, are also strongly 
correlated with the gaps
of the free neutron gas. For small average densities below $\rho \sim 0.02$
fm$^{-3}$, the $\Delta(R=0)$ values are larger than the corresponding gaps 
at the edge of the WS cell, as it can also be appreciated in the left panel
of the figure. However, for larger values of the average density in the WS cell,
this tendency is reversed and the gap at the edge is larger than at the origin
pointing to the increasing influence of the neutron gas.
These conclusions can also be drawn from an independent quantal   
study in Ref.\cite{grill11} where in the right panel of Fig. 4 very 
similar features
for the local gap values in different WS cells can be seen as in our TF study. 
We also show in Fig. 7, right panel, the LDA values of $\Delta(R=0)$ 
by the filled 
(green) triangles. We remark that those values undershoot quite strongly 
the corresponding TF values at the higher densities. 

For further illustration of this effect, we show in the two panels 
of Fig.~\ref{gapcrust2}
a comparison between  LDA and present TF results for the gaps in 
two particular WS cells. In the case of the largest cell whose 
representative nucleus is 
$^{500}_{40}$Zr, we see locally a huge difference in the surface region of 
the nucleus.
This simply stems from the fact that in this case the gap is very small and, 
therefore, the coherence length very large invalidating LDA. A study 
with examples a 
little less unfavorable for LDA is given in \cite{pas08} . This wrong 
behaviour of the local LDA gaps 
at low average densities can also be seen in Fig. 6 of Ref.~\cite{bal07a} . 
From that
figure we conclude that our TF calculation reproduces, qualitatively, 
the global trends of  the quantal gaps at low average densities.
The behaviour of the semiclassical gaps at high average density is clearly 
different
and it is dominated by the neutron gas as it can be seen in the right 
panel of
Fig.~\ref{gapcrust2} where the gap of the representative nucleus $^{1500}_{38}$Sr
is plotted as a function of the radius. Locally, LDA and TF show a depression 
in the center and the gap increases with increasing distance till it reaches 
its neutron
gas value. The central depression is stronger by about 30 percent in 
LDA than in TF. This 
behaviour is similar to the
one exhibited by the quantal gaps compared with the LDA ones displayed in
Ref.~\cite{cha10} .
In other words, this means that the contrast between inside and outside of the 
nucleus is much too pronounced in LDA, however, quantally as well as 
in TF this 
contrast is strongly attenuated by the proximity effect.
The semiclassical TF gaps at the Fermi level are displayed in the left panel 
of Fig.~\ref{gapcrust3}
as a function of chemical potential $\mu$. In the inner crust, i.e. for positive
values of $\mu$, they show a similar behaviour as the gaps at the 
edge of the WS cell
displayed in the right panel of Fig.~\ref{gapcrust1}. This behaviour can be 
expected as far as 
the gap at the Fermi level is rather an average quantity and, therefore, 
strongly
influenced by the neutron gas as it also happens in the HFB calculations 
of Ref.~\cite{pas11} .
In this figure we also include the gaps of some WS cells 
corresponding to 
the outer crust \
where all neutrons are bound with negative values of the chemical 
potential $\mu$. Again we see that the gap practically vanishes 
at zero chemical
potential when the neutrons start to drip.
In the right panel of Fig.~\ref{gapcrust3} we display the pairing energies 
per nucleon corresponding
to the WS cell of the inner crust. These energies are also correlated with the 
neutron gas and display
a similar behaviour as the one of the gaps at the Fermi energy. 
Again the pairing energy per
nucleon vanishes when neutrons arrive at the drip. It is rewarding that 
in the low average density regime
the TF pairing energies per nucleon follow the same trend as the quantal HFB 
ones\cite{grill11a} shown by the (red) filled triangles (however, a slightly 
different model with a somewhat stronger pairing force than in our case is 
used there).


For isolated nuclei at the neutron drip the situation may be somewhat different.
First, it may be that
in this situation the difference between HFB and BCS approaches is
more significant. Also strong shell fluctuations surely play an important role.
Somewhat conflicting results in this respect exist in the
literature. In ref \cite{ham05} very similar results to ours are found
for $S$-wave pairing, see Fig. 4 of this reference and also the discussion about 
it 
in \cite{Rom11} . On the other hand in \cite{taj05}
the gap seems to rise towards the drip before it bends down.
Similar results have also been found in \cite{hag11} .
The HFB calculation of Hamamoto has recently been repeated
and extended passing from negative chemical potentials to positive ones and
it was found that the $S$-wave gap clearly continues 
down to zero, touching zero at a slightly positive 
value of $\mu$ \cite{hagino} . 
In explaining why in other works the gap
is rising towards the drip, one has to keep in mind that an average gap should
be calculated with the pairing tensor and {\it not} with the density matrix.
The latter picks up the gaps at all energies which may not be small at all,
even though the gap at the Fermi level is very small, see the right panel of
 Fig.~\ref{slab1}. On the other
hand an average with the pairing tensor generally only picks up the (small) 
gaps around
the Fermi level. It also is intuitively clear that for other than
$S$-waves gaps the situation will be somewhat different. This is due to 
the finite centrifugal
barrier which keeps the wave function concentrated on the domain of the 
nucleus as
long as the corresponding energy stays below the barrier. However, large scale 
HFB calculations around the neutron drips of nuclei indicate that in general 
pairing is reduced at the drip line \cite{pas12}. In Fig. 10 we show a 
schematic study which may qualitatively reflect the real situation. 
A $Z,N$ dependent Woods Saxon potential (without spin orbit) given 
by Shlomo \cite{shlo92} was taken as the mean field and the BCS equation 
has been 
solved with the Gogny D1S pairing force \cite{D1S,D1Sa} .
Isotopic chains for two values 
of $Z$ have been claculated. For $Z=70$ the drip practically coincides with a 
shell closure of the neutrons and, therefore, the gap falls to zero at the 
drip for this case. On the other hand, for $Z=40$ the neutron drip does not 
coincide with a closed shell and then the gap has substantial values around 
the drip. Globally, however, a clear decreasing tendency of the gap towards 
the neutron drip can be observed as also reflected  by the TF values. 
The fact that neutron gaps decrease with increasing isospin was actually 
pointed out long time ago, see Refs.~\cite{lit05,yam09} . 
Real nuclei at the neutron drip may be either spherical or deformed (see 
Ref. \cite{Bruy}). For spherical drip nuclei it often happens that neutrons 
are at or very close to shell closure whereas for deformed nuclei this is not 
the case. The two situations then resemble the scenario displayed in the 
right and left panels of Fig.~\ref{Shlomogap} though, there, we imposed 
sphericity also for the case $Z$ = 40. 

A very promising possibility to recover shell effects is given by the fact 
that the latter mostly stem from the shell effects in the level density. 
We have discussed this problem with some detail in an earlier publication
\cite{vin11} . The basic idea is to replace 
in the TF gap equation (\ref{eq6}) the semiclassical 
$g(E)$ by its quantal 
counterpart, slightly smeared out by using gaussians centered at the 
quantal eigenvalues so that one obtains a continuous function 
of $E$ without destroying substantially the shell structure as is shown in 
Fig. \ref{D1Sgap}. Inserting this into the gap equation (\ref{eq6}), allows 
to recover almost completely the full 
quantal gaps as is shown in the right panel of Fig.~\ref{D1Sgap}.\\

In this work, we exclusively have treated pairing in BCS approximation. 
However, for certain situations and quantities, the more general HFB approach 
may be mandatory as, e.g., in the cases of rotation or a magnetic field. 
It is relatively straightforward to generalise the BCS-TF 
approach also to the HFB case. For this one has to consider fully non diagonal 
matrix elements $\langle n_1n_2|v|n_{1'}n_{2'}\rangle$. In the matrix 
elements, we have replaced $|\varphi_n(z)|^2$ by the TF expresion for 
the on shell 
density $\rho_E^{TF}(z)$. In the off diagonal pairing matrix element, we need 
wave functions and not densities. Therefore, in TF approximation, we then 
can use 
$\varphi_n(z) \rightarrow \sqrt{\rho_E^{TF}(z)}$ for the individual wave 
functions. Of course, this complicates the solution of the gap equation but 
this is always the price to pay when passing from BCS to HFB, quantally as 
well as semiclassically. The TF-HFB approach shall be investigated in future 
work \cite{sch12} . Let us finally mention that for spherical systems the TF 
approach can be generalised to partial waves as was done for the pairing matrix 
elements in \cite{vin03} .

\section{Summary}
Summarizing, we have studied superfluid fermions in a large container, either
external (cold atoms) or created self consistently (nuclei) for situations
where the top of the fluid reaches the edge of a small pocket located at the
origin of the wide confining potential. The gap drops to zero at the edge
before rising again when
the density fills up the outer container. This at first somewhat surprising
phenomenon can be explained quite straightforwardly. Such situations, as
already mentioned, can exist in cold atoms and
nuclei in the inner crust of neutron stars, two examples treated here with
their specific form of containers. For small systems, like isolated nuclei at
the neutron drip, the situation may be blurred by shell effects.

As an important second aspect of this work, we showed that a novel
Thomas-Fermi approach to
inhomogeneous situations can cope with situations where LDA
fails. This means that our TF approach is free of the restrictive condition,
prevailing for LDA, that the Cooper pair coherence length must be shorter
than a typical length $l$
(the oscillator length in the case of a harmonic container) over
which the mean field varies appreciably. On the contrary,
our TF theory has the usual TF validity criterion, namely that local
wavelengths must be shorter than $l$.

The accuracy
of our TF approach opens wide perspectives for a treatment of
inhomogeneous superfluid Fermi-systems with a great number of particles
not accessible for a quantal solution of the BCS (HFB) equations. Such
systems may be cold atoms in deformed containers (eventually reaching millions
of particles), superfluid-normal fluid (SN) interfaces, vortex profiles, etc.
As a matter of fact, as is well known \cite{RS} , the TF approach becomes the
more accurate, the larger the system. Thus the TF approximation is
complementary
to the quantal one in the sense that the former works where the latter is
difficult or even impossible to be obtained numerically.

Ideas and part of this paper have been published in earlier works, see for instance
Refs.~\cite{far00,prl} . A similar semiclassical approach also has been put 
forward  for mesoscopic systems in
Ref.~\cite{gar11} .

\vspace{0.3cm}


We thank K. Hagino for pointing to Ref.\cite{taj05} and sending  own
results prior to publication. Special thanks are due to A. Pastore and 
J. Margueron for valuable discussions and ongoing collaboration on the 
isolated nuclei problem \cite{pas12} .  
B. Juli\'a-D\'{\i}az  is greatfully acknowledged 
for providing us the quantal results of the double well 
potential.
This work has been partially supported by the IN2P3-CAICYT
collaboration (ACI-10-000592).
One of us (X.V.) acknowledges grants FIS2008-01661 and FIS2011-24154
(Spain and FEDER), 2009SGR-1289 (Spain) and Consolider Ingenio Programme 
CSD2007-00042 for financial support.

\vspace{0.3cm}



\begin{thebibliography}{9}
\bibitem{D1Sa}
The D1S force for pairing has been multiplied with
a factor 0.85 to compensate for the use of the bare mass in the calculation.
\end{thebibliography}
\end{document}